\documentclass[a4paper]{JHEP3} 

\usepackage{epsfig,multicol}
\usepackage{amsmath}
\usepackage{graphicx}
\graphicspath{{./}{figs/}}
\usepackage[latin1]{inputenc}
\usepackage{amssymb}
\usepackage{amsmath}
\usepackage{epsfig}
\usepackage{slashed}

\usepackage{feynarts} 

\usepackage{subfigure}
\usepackage{multirow}

\def\Z{\mathbb{Z}}

\def\1{\mathchoice{\rm 1\\\
mskip-4.2mu l}{\rm 1\mskip-4.2mu l}%
{\rm 1\mskip-4.6mu l}{\rm 1\mskip-5.2mu l}} 

\def\to{\rightarrow}

\def\Im{\mathrm{Im}}
\def\Re{\mathrm{Re}}

\newcommand\fverb{\setbox\pippobox=\hbox\bgroup\verb}
\newcommand\fverbdo{\egroup\medskip\noindent%
                      \fbox{\unhbox\pippobox}\ }

\newcommand\fverbit{\egroup\item[\fbox{\unhbox\pippobox}]}
\newbox\pippobox
\newcommand{\ie}{{\it i.e.}}
\newcommand{\eg}{{\it e.g.}}

\newcommand{\be}{\begin{equation}}
\newcommand{\ee}{\end{equation}}
\newcommand{\ba}{\begin{eqnarray}}
\newcommand{\ea}{\end{eqnarray}}
\newcommand{\bt}{\begin{tabular}}
\newcommand{\et}{\end{tabular}}
\newcommand{\bfig}{\begin{figure}}
\newcommand{\efig}{\end{figure}}

\title{Unconventional phenomenology of a minimal two-Higgs-doublet model}

\author{Simon de Visscher$^{*}$, Jean-Marc G\'{e}rard$^{*}$, Michel Herquet$^{\dagger}$, Vincent Lemaitre$^{*}$, Fabio Maltoni$^{*}$\\
* Universit\'{e} Catholique de Louvain, Chemin du Cyclotron 2, B-1348 Louvain-la-Neuve, Belgium\\
$\dagger$ Nikhef Theory Group, Science Park 105, 1098 XG Amsterdam, The Netherlands\\ 
E-mails: \email{simon.devisscher@uclouvain.be, jean-marc.gerard@uclouvain.be, mherquet@nikhef.nl, vincent.lemaitre@uclouvain.be, fabio.maltoni@uclouvain.be}

}

\preprint{NIKHEF-2009-003\\CP3-09-12}

\abstract{Two-Higgs-doublet models (2HDM) are simple extensions of the Standard Model (SM) where the scalar sector is enlarged by adding a weak doublet.  As a result, the Higgs potential depends in general on several free parameters which have to be carefully chosen to give predictions consistent with the current precision data. We consider a 2HDM invariant under a twisted custodial symmetry and depending only on three extra parameters beyond the SM ones. This model can naturally features an inverted mass spectrum with a light pseudoscalar state and a heavy SM-like Higgs boson. We thoroughly analyze direct and indirect constraints and present a few unconventional though promising signatures at the LHC.}


\begin{document}

\section{Introduction}
Even in the absence of any definitive experimental evidence, the mechanism of spontaneous symmetry breaking \cite{Higgs:1964pj,Englert:1964et,Guralnik:1964eu} remains the most promising candidate to consistently explain the existence of massive weak gauge bosons. Following Occam's principle, this mechanism is implemented within the Standard Model (SM) using a single $SU(2)_L$ Higgs doublet. Such a minimal solution displays three important advantages. First, it guarantees that a zero mass for the photon is not a mere accident \cite{Veltman:1997nm}. Second, it implies that the scalar potential is invariant under an $SU(2)_L\times SU(2)_R$ global symmetry, spontaneously broken down to a ``custodial'' $SU(2)_{L+R}$ \cite{Sikivie:1980hm}. This custodial symmetry naturally ensures that the one-loop quantum corrections to the successful tree-level relation
\begin{equation}
m_W^2=m_Z^2\left(\frac{g_L^2}{g_L^2+g_Y^2}\right)\,,
\label{eq:weakmass}
\end{equation}
where $g_L$ and $g_Y$ are the coupling constant associated with the gauge groups $SU(2)_L$ and $U(1)_Y$, vary at most logarithmically with the Higgs boson mass. Finally, it naturally explains the absence of tree-level Flavour Changing Neutral Currents (FCNC) \cite{Glashow:1976nt}.

Nevertheless, more complex possibilities involving additional and/or larger representations may \textit{a priori} be considered. Motivations for such enlarged Higgs sectors fall in general into two (not mutually exclusive) categories. Either they can be associated with extended symmetries, such as Supersymmetry and Grand Unification Theory (GUT) groups, or they can be justified by phenomenological arguments, such as the observed predominance of matter over anti-matter in our Universe.

The generic two-Higgs-doublet model, where the SM scalar sector is enlarged with only one additional $SU(2)_L$ doublet, appears as an attractive laboratory. Though relatively simple, it can display a rich phenomenology. Specific realisations of this model can also be predicted, or required, by Beyond the Standard Model (BSM) theories such as the Minimal Supersymmetric SM (MSSM) or $SO(10)$-invariant GUTs. However, contrary to the SM scalar sector, the most general 2HDM Higgs potential does not guarantee $m_\gamma=0$, neither is invariant under custodial symmetry nor under $CP$ symmetry, and leads to large FCNC. In order to naturally explain the absence of large corrections to precisely measured quantities, additional symmetries in the Higgs sector of the theory must therefore be imposed. In particular, the existence of a custodial symmetry, even if only approximate, is a highly desirable feature to explain the smallness of the Higgs boson one-loop correction to relation \eqref{eq:weakmass}. While the \textit{logarithmic} contribution of the SM Higgs boson with a mass around the electroweak scale is compatible with all current precision data, any additional \textit{quadratic} contribution is severely constrained. 

In Ref.~\cite{Gerard:2007kn}, a general expression for the $SU(2)_L\times SU(2)_R$ chiral symmetry playing this important ``custodial'' role was presented. A phase arbitrariness in the definition of this symmetry has also been emphasised, in conjunction with the generic definition of the $CP$ symmetry. As a result \textit{two} physically different custodial limits for a $CP$ conserving Higgs potential were found, characterised by the mass degeneracy of the charged Higgs pair $H^\pm$ with either a pseudoscalar state $A^0$ or a scalar state $H^0$.  

The present work describes the rather unconventional phenomenology associated with the second implementation of the custodial symmetry, which we refer to as ``twisted''. As already mentioned in Ref.~\cite{Gerard:2007kn}, this scenario gives rise to experimental signatures that are qualitatively very different from those usually considered in models like, for example, the MSSM. The main difference is the possibility for the pseudoscalar $A^0$ to be as light as a few tens of GeV with the scalar $h^0$ heavier than the SM expectation. This opens new perspectives for unusual decays like $h^0\to A^0 A^0, H^0 H^0, H^+ H^-$, $H^\pm\to W^\pm A^0$ and $H^0\to Z A^0$ that can become dominant and drastically affect our discovery strategies. 
Note that some of these decay modes may also appear in several BSM contexts such as a small $\tan\beta$ MSSM scenario (see Ref.~\cite{Dermisek:2008id}), the Next to Minimal Supersymmetric SM (NMSSM, see Ref. \cite{Dermisek:2008uu}) or the  $CP$ violating MSSM. However, current analyses of the associated signals appear to be limited to very specific regions of the parameter space.

The paper is organised as follows. First, we implement a minimal version of the twisted scenario proposed in Ref.~\cite{Gerard:2007kn} and discuss the possible Yukawa couplings. Second, various theoretical, indirect and direct constraints are examined in order to restrict the model parameter space and to define a pertinent set of benchmark points. Finally, production and decay modes are systematically considered and a few interesting experimental signatures identified. Three particular channels that clearly illustrate the LHC discovery potential in this scenario are then considered in more detail.

\section{A minimal model}
\label{sec:model}
Consider a 2HDM based on the $SU(2)_L$ doublets $\phi_1$ and $\phi_2$ with the same hypercharge $Y=+1$. If one imposes only gauge invariance, the most general renormalisable scalar potential reads, in standard notation,
\begin{eqnarray}
 V( \phi_1, \phi_2)&=&-m_1^2 \phi_1^\dag \phi_1 -m_2 ^2\phi_2^\dag \phi_2-(m_3^2 \phi_1^\dag \phi_2+\mathrm{h.c.})+\frac{\lambda_1}{2}( \phi_1^\dag \phi_1)^2\nonumber\\
& &+\frac{\lambda_2}{2}( \phi_2^\dag \phi_2)^2+\lambda_3( \phi_1^\dag \phi_1)( \phi_2^\dag \phi_2)+\lambda_4( \phi_1^\dag \phi_2)( \phi_2^\dag \phi_1)\nonumber\\
& &+\left[\frac{\lambda_5}{2}( \phi_1^\dag \phi_2)^2+\left(\lambda_6( \phi_1^\dag \phi_1)+\lambda_7( \phi_2^\dag \phi_2)\right)( \phi_1^\dag \phi_2)+\mathrm{h.c.}\right]\,,
\label{eq:genpot}
\end{eqnarray}
where $m_{1,2}^2$ and $\lambda_{1,2,3,4}$ are real parameters while $m_3^2$ and $\lambda_{5,6,7}$ are \textit{a priori} complex. 

This potential, though very general, has two important drawbacks. First, it contains a large number of free parameters that challenges any complete in-depth analysis of its phenomenological consequences. Second, in the absence of additional global symmetries like those accidentally preserved by the Standard Model scalar sector, large corrections to various tightly constrained precision parameters are expected. Our aim is to address both issues by restricting \eqref{eq:genpot} by means of well-motivated global symmetries, yet leaving enough room for unconventional experimental signatures.

Let us first assume that the electromagnetic gauge symmetry is preserved,\ie , that the vacuum expectation values (v.e.v.) of $\phi_1$ and $\phi_2$ are aligned in the $SU(2)_L$ space in such a way that a single $SU(2)_L$ gauge transformation suffices to rotate them to the neutral components.\footnote{This phenomenologically motivated assumption is in fact rigorously justified in the context of the restricted models to be considered in the following (e.g., see \cite{DiazCruz:1992uw,Ferreira:2004yd}).} After a suitable $SU(2)\times U(1)_Y$ transformation, one has
\begin{equation}
\label{eq:2hdmvev}
\langle\phi_1\rangle = \frac{1}{\sqrt{2}}\left(
\begin{array}{c}
  0\\
  v_1
\end{array}
\right)\quad\mathrm{and}\quad 
\langle\phi_2\rangle = \frac{1}{\sqrt{2}}\left(
\begin{array}{c}
  0\\
  v_2 e^{i\theta}
\end{array}
\right)
\end{equation}
with $v_1$ and $v_2$ two real parameters such that $v_1^2+v_2^2\equiv v^2=(\sqrt{2} G_F)^{-1}$ and $v_2/v_1\equiv \tan\beta$. An important feature of the 2HDM is the freedom to redefine the two scalar fields $\phi_1$ and $\phi_2$ using arbitrary $U(2)$ transformations acting in the ``flavour'' space,\ie ,
\begin{equation}
\label{eq:unit2HDM}
\left(
\begin{array}{c}
  \phi_1\\
  \phi_2
\end{array}
\right)\rightarrow
\left(
\begin{array}{c}
  \phi_1'\\
  \phi_2'
\end{array}
\right)\equiv U\left(
\begin{array}{c}
  \phi_1\\
  \phi_2
\end{array}
\right)
\quad,\quad U\in U(2)\quad.
\end{equation}
Transformation \eqref{eq:unit2HDM} leaves the canonically normalized, gauge-covariant kinetic energy terms invariant. This notion of \textit{basis invariance} has been emphasized in \cite{Branco:1999fs} and considered in great detail more recently in \cite{Ginzburg:2004vp} and \cite{Davidson:2005cw,Haber:2006ue}. From now on, let us take advantage of this property to select one of the \textit{Higgs basis} where only one of the two Higgs field acquires a non-zero v.e.v.:
\begin{equation}
\label{eq:HB}
v_1=v\quad\mathrm{and}\quad v_2=0\,.
\end{equation}
Note that the Higgs basis is not univocally defined since the reparametrization $\phi_2\to e^{i\alpha} \phi_2$ leaves the condition \eqref{eq:HB} invariant. As illustrated in the following, this (set of) basis is particularly convenient since $\phi_1$ contains all the SM would-be Goldstone fields, such that any generalization of SM global symmetries is almost straightforward. 

In the Higgs basis, the most general definition of the $CP$ symmetry simply reads (e.g., see \cite{Branco:1999fs}):
\begin{eqnarray}
(CP)\phi_1(t,\vec{x})(CP)^\dag&=&  {\phi_1}^*(t,-\vec{x})\nonumber\\
(CP)\phi_2(t,\vec{x})(CP)^\dag&=& e^{i\delta} {\phi_2}^*(t,-\vec{x})
\label{eq:cp2HDMgen}
\end{eqnarray}
and displays a single arbitrary phase. For the sake of simplicity, and without any loss of generality, we take advantage of the $\phi_2$ phase freedom to conventionally fix the value of the $CP$ phase at one, \ie , $\delta=0$. If one imposes $CP$ invariance on the bosonic sector of the theory, then relations \eqref{eq:cp2HDMgen} restrict all the scalar potential parameters to be real. The resulting potential is thus described by only ten free parameters, \ie , four less than the full one given in Eq.~\eqref{eq:genpot} since $m_3^2$ and $\lambda_{5,6,7}$ are now taken real. This restricted potential is denoted $V^{CP}_{10}$ in the following. 

In the Higgs basis, the most general custodial symmetry transformation for a two-Higgs-doublet model also takes the simple form \cite{Gerard:2007kn}:
\begin{equation}
\label{eq:custtransf}
M_1\rightarrow U_L M_1 U_R^\dag\quad,\quad M_2\rightarrow U_L M_2 V_R^\dag
\end{equation}
where $M_i$ corresponds to the usual $[1/2,1/2]$ representation of $\phi_i$, i.e., 
\begin{equation}
M_i\equiv\left(
\begin{array}{cc}
\phi_i^0 & \phi_i^+ \\
-(\phi_i^+)^* & (\phi_i^0)^*
\end{array}\right)\,.
\end{equation}
The right transformation $V_R$ for $M_2$ in \eqref{eq:custtransf} is not necessarily fixed by the right transformation $U_R$ for $M_1$. Only $SU(2)_L\times U(1)_Y$ is a local symmetry of the Lagrangian.  So, we still have the freedom to choose $V_R=X^\dag U_R X$ if the two-by-two unitary matrix $X$ commutes with $\exp(iT_3^R)$, where $T_3^R$ is the diagonal generator of the global $SU(2)_R$. We thus fix $X\equiv \exp(i\gamma T_3^R)$ where $\gamma$ is an arbitrary angle. 

Now the crucial point is that we have already used the reparametrization invariance of $\phi_2$ to fix the $CP$-phase. As an all-important consequence, the custodial angle $\gamma$ is a truthfully free parameter. Depending on its value, invariance of the general $CP$ invariant potential $V^{CP}_{10}$ under the custodial symmetry can be obtained for different limits of the potential parameters:
\begin{enumerate}
\item If $\gamma = 0$, the $CP$-even combinations $\phi_1^\dag\phi_1$, $\phi_2^\dag\phi_2$ and $(\phi_1^\dag\phi_2+\phi_2^\dag\phi_1)$ are invariant under \eqref{eq:custtransf}. Invariance of the $CP$ conserving potential  $V^{CP}_{10}$ under this custodial symmetry is restored in the limit $\lambda_4=\lambda_5$ such that
\begin{eqnarray}
 V^{\mathrm{usual}}_9( \phi_1, \phi_2)&=&-m_1^2 \phi_1^\dag \phi_1 -m_2 ^2\phi_2^\dag \phi_2-m_3^2(\phi_1^\dag\phi_2+\phi_2^\dag\phi_1)+\frac{\lambda_1}{2}( \phi_1^\dag \phi_1)^2\nonumber\\
& &+\frac{\lambda_2}{2}( \phi_2^\dag \phi_2)^2+\lambda_3( \phi_1^\dag \phi_1)( \phi_2^\dag \phi_2)+\frac{\lambda_5}{2}(\phi_1^\dag\phi_2+\phi_2^\dag\phi_1)^2\nonumber\\
& &+(\phi_1^\dag\phi_2+\phi_2^\dag\phi_1)\left[\lambda_6( \phi_1^\dag \phi_1)+\lambda_7( \phi_2^\dag \phi_2)\right].
\label{eq:usualpot}
\end{eqnarray}
This gives the usual realization of the custodial symmetry in the 2HDM.
\item If $\gamma = \pi$, the $CP$-even combinations $\phi_1^\dag\phi_1$, $\phi_2^\dag\phi_2$ and the $CP$-odd combination $(\phi_1^\dag\phi_2-\phi_2^\dag\phi_1)$ are invariant under \eqref{eq:custtransf}. Invariance of the potential $V^{CP}_{10}$ under this ``twisted'' custodial symmetry is restored in the limit $\lambda_4=-\lambda_5$ and $m_3^2=\lambda_6=\lambda_7=0$, giving
\begin{eqnarray}
 V^{\mathrm{twisted}}_6( \phi_1, \phi_2)&=&-m_1^2 \phi_1^\dag \phi_1 -m_2 ^2\phi_2^\dag \phi_2+\frac{\lambda_1}{2}( \phi_1^\dag \phi_1)^2+\frac{\lambda_2}{2}( \phi_2^\dag \phi_2)^2\nonumber\\
& &+\lambda_3( \phi_1^\dag \phi_1)( \phi_2^\dag \phi_2)+\frac{\lambda_5}{2}(\phi_1^\dag\phi_2-\phi_2^\dag\phi_1)^2.
\label{eq:twistedpot}
\end{eqnarray}
\item If $\gamma\neq 0$ and $\pi$, the only combinations of $\phi_1$ and $\phi_2$ that are, at the same time, $CP$-eigenstates and invariant under \eqref{eq:custtransf} are $\phi_1^\dag\phi_1$ and $\phi_2^\dag\phi_2$. The resulting, highly symmetric potential reads then
\begin{eqnarray}
 V^{\mathrm{sym}}_5( \phi_1, \phi_2)&=&-m_1^2 \phi_1^\dag \phi_1 -m_2 ^2\phi_2^\dag \phi_2+\frac{\lambda_1}{2}( \phi_1^\dag \phi_1)^2+\frac{\lambda_2}{2}( \phi_2^\dag \phi_2)^2\nonumber\\
 & &+\lambda_3( \phi_1^\dag \phi_1)( \phi_2^\dag \phi_2)\,.
\label{eq:symdpot}
\end{eqnarray}
In fact this potential is so constrained that the $SU(2)_L$ representation of $\phi_2$ is not even fixed by the interaction term and any representation (singlet or higher) is allowed. 
\end{enumerate}
Out of the three cases above, only the second one is novel and from now on, we focus exclusively on it. This ``twisted'' scenario, characterized by a relatively low number of free parameters (at least compared to the usual case), yet displays the possibility of interesting phenomenology due to the non-trivial mixed term $(\phi_1^\dag\phi_2-\phi_2^\dag\phi_1)^2$ in Eq.~\eqref{eq:twistedpot}.

The last symmetry to be considered in the twisted scenario is the $\Z_2$ symmetry
\begin{equation}
\phi_1 \to \phi_1 \quad,\quad \phi_2\to -\phi_2\,,
\label{eq:z2transfo}
\end{equation}
known (assuming suitable $\Z_2$-charges for the SM fermions) to be a simple, yet elegant, way to suppress tree-level FCNC in any 2HDM  \cite{Glashow:1976nt}. The potential  \eqref{eq:twistedpot} is naturally invariant under this symmetry without further assumptions. However, if such a $\Z_2$ symmetry is only manifest in a Higgs basis, all fermions are forced to be $\Z_2$-even in order to couple to $\phi_1$ and get non vanishing mass terms. If such is the case, they cannot couple to $\phi_2$. This provides a natural frame for dark matter (e.g., see \cite{LopezHonorez:2006gr}) restricting, however, considerably the possible phenomenological signatures at colliders. 

This apparent limitation can be circumvented if the $SO(2)$ rotation of angle $\beta=\arctan(v_2/v_1)$ required to go from a generic basis (where both v.e.v. are real and non zero) to the Higgs basis \eqref{eq:HB} is promoted to a symmetry of the potential. In this case, a $\Z_2$ symmetry defined in the Higgs basis would remain manifest in any related basis, and \textit{vice versa}. Since the only matrices to commute with the generator of the $SO(2)$ symmetry, \ie , the second Pauli matrix $\tau_2$, are the identity matrix and $\tau_2$ itself, the two invariants are $(\phi_1^\dag \phi_1+ \phi_2^\dag \phi_2)$ and $(\phi_1^\dag\phi_2-\phi_2^\dag\phi_1)$. Imposing invariance of the quartic part of potential \eqref{eq:twistedpot}  under this $SO(2)$ (a softly-broken $\Z_2$ symmetry being sufficient to ensure the absence of large FCNC effects) reduces the total number of parameters from six to four:
\begin{eqnarray}
 V^{\mathrm{min.}}_4( \phi_1, \phi_2)&=&-m_1^2 \phi_1^\dag \phi_1 -m_2 ^2\phi_2^\dag \phi_2+\frac{\lambda_S}{2}( \phi_1^\dag \phi_1+ \phi_2^\dag \phi_2)^2\nonumber\\
 & &+\frac{\lambda_{AS}}{2}(\phi_1^\dag\phi_2-\phi_2^\dag\phi_1)^2\,.
\label{eq:twistedZ2pot}
\end{eqnarray}
Our forthcoming study of unconventional phenomenology is based on this minimal potential where, for convenience, we have introduced the symmetric ($\lambda_S=\lambda_1=\lambda_2=\lambda_3$) and antisymmetric ($\lambda_{AS}=\lambda_5=-\lambda_4$) notation for the coefficients of the corresponding quartic terms.

It is straightforward to determine the physical spectrum associated with the potential \eqref{eq:twistedZ2pot} in the Higgs basis. First, it contains a $CP$-even SM-like Higgs boson $h^0\equiv \sqrt{2}(\Re(\phi_1)-v/\sqrt{2})$ with squared mass
\begin{equation}
m_{h^0}^2=\lambda_S v^2=2m_1^2\,.
\label{eq:h0massTC}
\end{equation}
Second, it displays a pair of charged Higgs bosons and a $CP$-even scalar $H^0\equiv -\sqrt{2}\Re(\phi_2)$, forming a triplet under the custodial symmetry with
\begin{equation}
m_T^2\equiv m_{H^\pm}^2=m_{H^0}^2=m_1^2-m_2^2\,.
\label{eq:masstriplet}
\end{equation}
Last but not least, it allows the pseudoscalar state $A^0=\sqrt{2}\Im(\phi_2)$ to be singlet under the twisted custodial symmetry with
\begin{equation}
m_{A^0}^2=m_{T}^2-\lambda_{AS}v^2\,.
\label{eq:A0H0massTC}
\end{equation}
Since $h^0$ remains the only massive Higgs boson in the limit of an exact $SO(8)$ symmetry acting on the real doublet components ($\lambda_{AS}\to 0$, $m_1\to m_2$), one may thus expect the following inverted hierarchy: 
\begin{equation}
m_{A^0} < m_{H^0,H^\pm} < m_{h^0}\quad\mathrm{if}\quad\lambda_{AS}>0\,.
\label{eq:invspec}
\end{equation}

As seen from Eq. \eqref{eq:masstriplet}, the main consequence of the twisted custodial symmetry is a mass degeneracy between the pair of charged Higgs bosons $H^\pm$ and the \textit{scalar} state $H^0$. This contrasts with the usual approach that links the custodial symmetry to a mass degeneracy between the pair of charged Higgs bosons $H^\pm$ and the \textit{pseudoscalar} state $A^0$. The existence of a genuine twisted scenario in the 2HDM, and more generically the interpretation of the possible interplay between the custodial and $CP$ symmetries in terms of 
aligned/misaligned  phases, Eqs.~\eqref{eq:usualpot}/\eqref{eq:twistedpot}, is a non trivial observation. Previous phenomenological studies of the $\rho\approx 1$ constraint had already noted a vanishing contribution in the $m_{H^\pm}^2\approx m_{H^0}^2$ limit (\cite{Toussaint:1978zm} or more recently \cite{Chankowski:2000an} and \cite{Barbieri:2006dq}) but none of them did interpret this property in terms of symmetries. The interplay between the $CP$ and custodial symmetry in 2HDM has also been considered in \cite{Pomarol:1993mu}, but the second $CP$ conserving custodial scenario discussed there is in fact physically indistinguishable from the ``usual'' case since $H^0$ can be redefined as $CP$ odd. Here, the usual and twisted cases clearly lead to different physical situations, as will be shown explicitly by considering, for example, quantum corrections to the $T$ parameter (see Fig.~\ref{fig:deltaT}) or to the muon anomalous moment (Sec.~\ref{subsec:gminus2}).

Regarding Yukawa couplings, the presence of a softly-broken $\Z_2$ symmetry allows to define type I and type II models \cite{Barger:1989fj}, with a very different phenomenology. Type I models correspond to the case where only one Higgs doublet couples to all fermions in the generic basis. This can be enforced by assuming that all left and right handed fermions are even under the softly-broken $\Z_2$ symmetry \eqref{eq:z2transfo}. The Yukawa interaction terms in the Higgs basis then read
\begin{eqnarray}
\mathcal{L}_Y^I&=&- \frac{\sqrt{2}}{v}\overline{Q_L} M_d (\phi_1-\tan\beta\;\phi_2) d_R  - \frac{\sqrt{2}}{v}\overline{Q_L} M_u (\tilde{\phi_1}-\tan\beta\;\tilde{\phi_2})  u_R +\mathrm{h. c.}\,,
\label{eq:yuklag2HDMI}
\end{eqnarray}
with $\tilde{\phi_1}= i\tau_2 \phi_1^*$. In terms of physical states, the couplings of $h^0$ to fermions are identical to those observed in the SM, {\it  i.e.}, directly proportional to the fermion mass matrix $M$, while those of $H^0$, $A^0$ and $H^\pm$ are simply rescaled by a factor\footnote{Depending on which doublet is conventionally chosen to couple to all fermions in the generic basis, this scale factor can also be $\cot\beta$ as often considered in the literature.} $\tan\beta$. If $\tan\beta\to 0$, $\phi_2$ decouples from the fermionic sector and provides a viable dark matter candidate \cite{LopezHonorez:2006gr}. 

In type II models, one of the two Higgs doublet couples to the down type quarks and to the charged leptons, while the other one only couples to the up type quarks in the generic basis. This is easily achieved by assuming all fermionic fields to be even under the softly-broken $\Z_2$ symmetry \eqref{eq:z2transfo} except for the right handed up type quarks fields which are odd.  The Yukawa interaction terms in the Higgs basis now read
\begin{eqnarray}
\mathcal{L}_Y^{II}&=&- \frac{\sqrt{2}}{v}\overline{Q_L} M_d (\phi_1-\tan\beta\;\phi_2) d_R  - \frac{\sqrt{2}}{v}\overline{Q_L} M_u (\tilde{\phi_1}+\cot\beta\;\tilde{\phi_2}) u_R +\mathrm{h. c.}\,.
\label{eq:yuklag2HDMII}
\end{eqnarray}
In terms of physical states, the couplings of $h^0$ to fermions are still identical to those observed in the SM, but those of $H^0$, $A^0$ and $H^\pm$ are now rescaled by a factor $\tan\beta$ or $\cot\beta$ depending upon whether the fermionic current involves down or up type fermions.

The minimal two-Higgs-doublet model (M2HDM) considered in this work is defined by Eqs. \eqref{eq:twistedZ2pot}, \eqref{eq:yuklag2HDMI} and \eqref{eq:yuklag2HDMII}. It has only four free parameters: $m_{h^0}$,  $m_T$ and $m_{A^0}$ for the scalar potential, and $\tan\beta$ for the Yukawa interactions. Inspired at first by a twisted custodial symmetry, one can also find a generic 2HDM whose parameters in the potential are adjusted in such a way to implement the inverted mass spectrum~\eqref{eq:invspec}.

\section{Parameter constraints}
\label{sec:constr}
In order to reduce the parameter space, let us now review the theoretical, indirect and direct constraints at our disposal.

\subsection{Theoretical constraints}
The first type of relevant theoretical constraints are the vacuum stability conditions. They come from the requirement of a positive potential for large classical values of the fields in any direction in the $(\phi_1,\phi_2)$ plane. These constraints can be obtained by considering only the quartic terms of the potential \cite{Deshpande:1977rw,Ginzburg:2004vp}. In the context of the minimal potential defined in Eq. \eqref{eq:twistedZ2pot}, they read
\begin{equation}
\lambda_S>0\quad\textrm{and}\quad
\lambda_S>\lambda_{AS}\,,
\label{eq:stability}
\end{equation}
or equivalently, in terms of the physical masses,
\begin{equation}
m_{h^0}^2>0\quad\mathrm{and}\quad m_{h^0}^2>m_T^2-m_{A^0}^2\,.
\label{eq:hierarchyscal}
\end{equation}
In fact, the squared mass of all the (pseudo)scalars are, by definition, positive if the v.e.v. correspond to a minimum of the potential \eqref{eq:twistedZ2pot}.

\begin{figure}[htbp]
\begin{center}
\subfigure[Unitarity]{\includegraphics[width=0.45\textwidth]{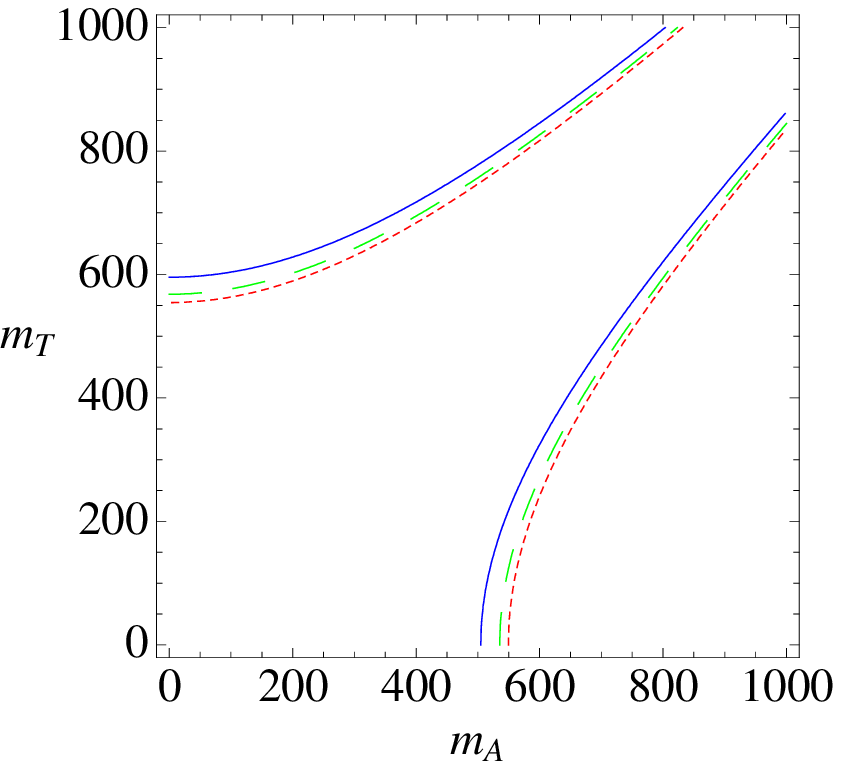}}
\subfigure[Perturbativity]{\includegraphics[width=0.45\textwidth]{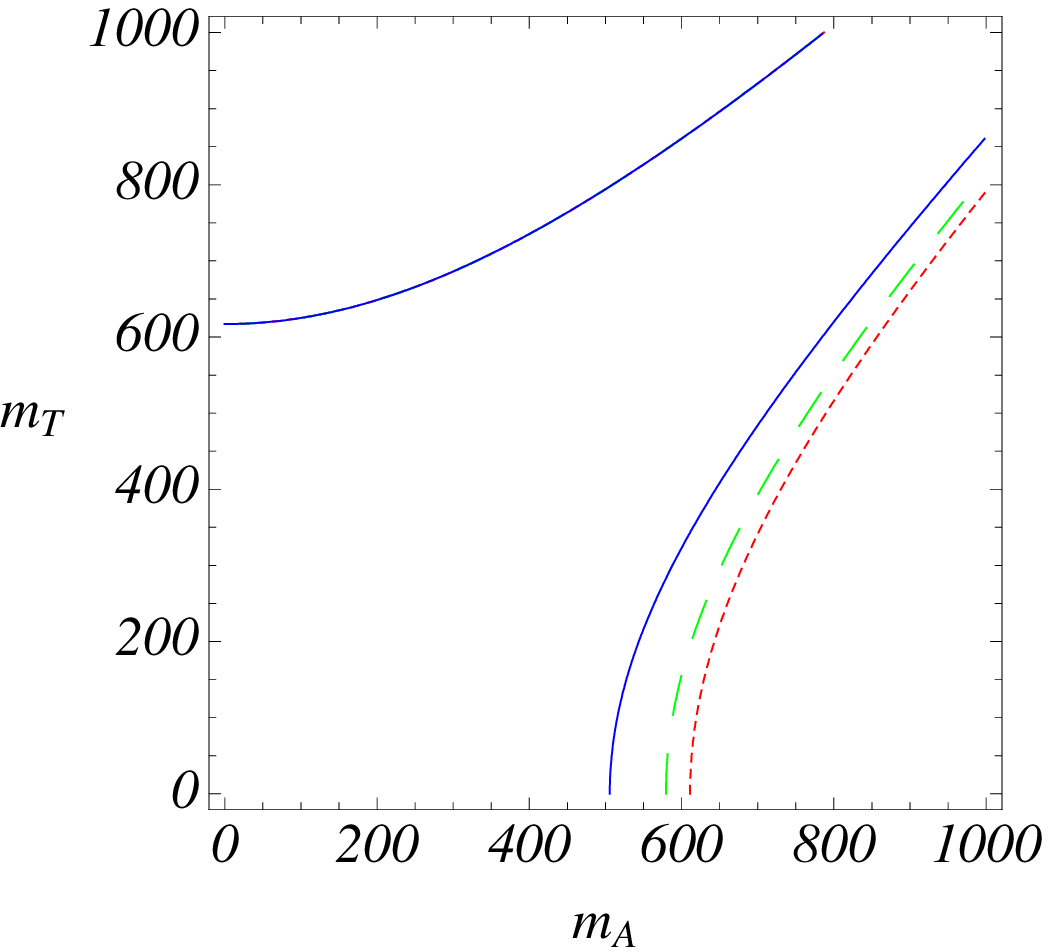}}
\caption{Unitarity and perturbativity constraints in the $(m_{A^0},m_T)$ plane for the M2HDM. Dotted red lines are limits for $m_{h^0}=120$ GeV, dashed green lines for $m_{h^0}=300$ GeV and plain blue lines for $m_{h^0}=500$ GeV. The allowed regions lie between these lines.}
\label{fig:unittwisted}
\end{center}
\end{figure}

The second set of relevant constraints comes from the requirement of unitarity for all the possible scattering processes involving the new scalar particles. They have been worked out for both $CP$ conserving and $CP$ violating potentials \cite{Akeroyd:2000wc,Ginzburg:2005dt} and can be advantageously summarized as
\begin{equation}
|\Lambda_{YI^3}^{\Z_2}|<8\pi\,,
\label{eq:uniTC}
\end{equation}
where $|\Lambda_{YI^3}^{\Z_2}|$ are the eigenvalues of the high energy scattering matrix for different quantum numbers of the initial state: total hypercharge $Y$, weak isospin $I^3$ and $\Z_2$ parity. In the M2HDM (see Eq.~\eqref{eq:twistedZ2pot}), the relevant contributions are
\begin{eqnarray}
\Lambda_{21}^{\mathrm{even}}&=&\lambda_S\pm \lambda_{AS}\nonumber\\
\Lambda_{00}^{\mathrm{even}}&=&3\lambda_S\pm(2\lambda_S-\lambda_{AS})\nonumber\\
\Lambda_{00}^{\mathrm{odd}}&=&\lambda_S-2\lambda_{AS}\pm 3\lambda_{AS}\,.
\end{eqnarray}
The two sign possibilities appearing in these expressions correspond to the scattering of different initial states with the same quantum numbers. Using relations \eqref{eq:h0massTC} and \eqref{eq:A0H0massTC}, constraints \eqref{eq:uniTC} restrict the possible values of the scalar masses. In particular, in the limit where all scalar masses are small except for one, say $m_S$, one has the upper bound
\begin{equation}
m_{S}\lesssim 550\,\mathrm{GeV}\,.
\end{equation}
When the Higgs masses are non negligible, the unitarity requirement may help to restrict, for example, the allowed region in the $(m_{A^0},m_T)$ plane for given values of $m_{h^0}$, as displayed in Fig.~\ref{fig:unittwisted}(a).
A naive estimate of the region compatible with the perturbative approach is shown in Fig.~\ref{fig:unittwisted}(b). In this region, all three- and four- scalar vertices are bounded by $4\pi$, such that the effective parameter of perturbation theory is smaller than one. Let us emphasize that more sophisticated treatments of the perturbativity constraint applied on similar models (SM and MSSM) only lead to minor corrections with respect to this naive approximation  (e.g., see  \cite{Tobe:2002zj}).

\subsection{Precision measurement constraints}
\subsubsection{Electroweak precision parameters}
The total contribution of the new scalar states to the $T$ parameter \cite{Peskin:1991sw} in the context of the multi-Higgs-doublet model has been computed recently in Ref.~\cite{Grimus:2007if}. Here, we focus on the correction $\Delta T$ from the $CP$ conserving potential $V_{10}^{CP}$. Those are well-known in the limit where all scalar squared masses are larger than $m_Z^2$ \cite{Toussaint:1978zm,Lytel:1980zh}. If one of the scalar is lighter than $m_Z$, however, the exact expression obtained in Ref.~\cite{Haber:1993wf} and reported in Ref.~\cite{He:2001tp} is more suited.

\begin{figure}[htbp]
\begin{center}
\includegraphics[width=9cm]{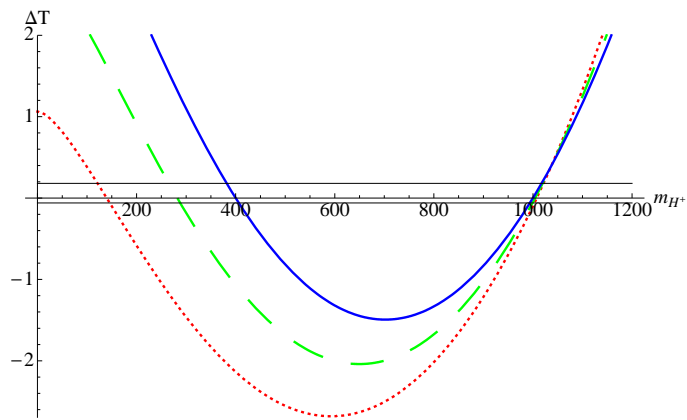}
\caption{The $\Delta T$ correction with respect to the charged Higgs pair mass (in GeV) from the $CP$ conserving potential in a 2HDM, for $(\beta-\alpha)=0$ (dotted red), $\pi/4$ (dashed green) and $\pi/2$ (solid blue). The other scalar masses are fixed to $m_{h^0}=150$ GeV, $m_{H^0}=400$ GeV and $m_{A^0}=1$ TeV. The thin horizontal black lines shows the 2$\sigma$ experimental limits on $\Delta T$ (assuming $\Delta S=0$) \cite{lepewwg}.}
\label{fig:deltaT}
\end{center}
\end{figure}

The results for $\Delta T$ as a function of $m_{H^\pm}$ are shown in Fig.~\ref{fig:deltaT}. With different values of $(\beta-\alpha)$, where $\alpha$ is the mixing angle between the two $CP$-even eigenstates in $V_{10}^{CP}$ and fixed values for all other scalar masses,
Fig.~\ref{fig:deltaT} displays two cases where $\Delta T$ is close to zero:
\begin{enumerate}
\item The solution $m_{H^\pm}\approx m_{A^0}$, independently of the value of $(\beta-\alpha)$ and the masses of the other scalars;
\item A continuum of solutions ranging from $m_{H^\pm}\approx m_{H^0}$ when $\beta-\alpha=\pi/2$, to $m_{H^\pm}\approx m_{h^0}$ when $\beta-\alpha=0$.
\end{enumerate}
The first possibility corresponds to the usual custodial scenario leading to the potential \eqref{eq:usualpot}, while the second one corresponds to the ``twisted'' case with the potential \eqref{eq:twistedpot}. In fact, the two solutions $m_{H^\pm}=m_{h^0}$ and $m_{H^\pm}=m_{H^0}$ in the second case are related through a $\pi/2$ shift in $\alpha$,{\it  i.e.}, a renaming of $h^0$ and $H^0$. The continuum of solutions between these two mass degeneracies corresponds to the situation where the neutral state  belonging to the custodial triplet is not a mass eigenstate, but a mixture of $h^0$ and $H^0$. 

\begin{figure}[htbp]
\begin{center}
\includegraphics[width=9cm]{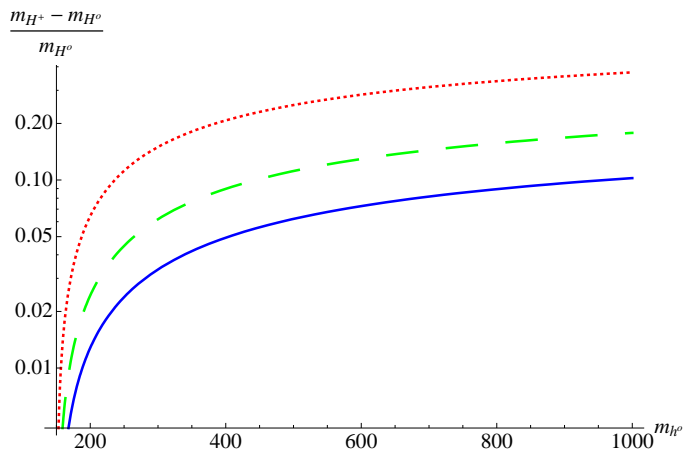}
\caption{Soft relative corrections to $m_{H^\pm}=m_{H^0}$ in the M2HDM that give $\Delta T=0$ as a function of the SM-like Higgs boson mass $m_{h^0}$ (in GeV). The dotted red, dashed green and solid blue lines correspond  to $m_{H^0}$=200, 300 and 400 GeV, respectively. The $A^0$ mass is fixed at 100 GeV but does not affect the results if $m_{A^0}\ll m_{h^0},m_T$.}
\label{fig:triplet_comp}
\end{center}
\end{figure}

An interesting possibility in the framework of the M2HDM based on potential \eqref{eq:twistedZ2pot} arises when the pseudoscalar state $A^0$ is light while all the other scalars are heavier ($> 100$ GeV). In this case, a moderate deviation from the degeneracy $m_{H^\pm}=m_{H^0}$ could compensate the large logarithmic contributions involving $m_{h^0}$. This can be seen directly from the first order analytical approximation for $\Delta T$ in the $m_{H^\pm}\approx m_{H^0}$ region \cite{Chankowski:2000an}:
\begin{eqnarray}
\Delta T&\approx& \frac{1}{16\pi m_W^2 \cos^2\theta_W}\times\left\{\cot^2\theta_W\left(\frac{m^2_{H^\pm}-m^2_{H^0}}{2}\right)\right.\nonumber\\
& &\left.-3m^2_W\left(\log\frac{m^2_{h^0}}{m^2_W}+\frac{1}{\sin^2\theta_W}\log\frac{m^2_W}{m^2_Z}+\frac{1}{6}\right)\right\}\quad.
\end{eqnarray}
The amount of breaking required for a vanishing $T$ as a function of the $h^0$ Higgs mass is shown in Fig.~\ref{fig:triplet_comp}.
Using the effective potential approach \cite{Coleman:1973jx}, we estimated the one-loop corrections to the tree-level relations \eqref{eq:masstriplet} in the context of the M2HDM. Non-vanishing contributions can come only from terms that explicitly break the custodial symmetry, \ie, the gauge and Yukawa interactions. The main contribution arising from the top (and bottom) quark loops leads to a relative mass difference $(m_{H^+}-m_{H^0})/m_{H^0}$ smaller than 1\% for both type I and type II models. The remaining mass difference required to satisfy the $T$ parameter constraint (see Fig.~\ref{fig:triplet_comp}) should thus either be explained by the presence of a small custodial breaking term in the scalar potential, like for example $\epsilon(\phi_1^\dag\phi_2+\phi_2^\dag\phi_1)^2$, or by the presence of other BSM contributions (\eg, heavy fields leading to additional radiative corrections). Yet, those effects would have a negligible impact on the phenomenology beyond $\Delta T$.

Contrary to the $T$ parameter, the $S$ and $U$ parameters only depend logarithmically on the mass of the new scalar particles. The exact one-loop contribution from $V_{10}^{CP}$ has been evaluated in Refs. \cite{Haber:1993wf,He:2001tp}. The numerical results for these contributions show that scenarios with a light pseudoscalar and a heavy degenerate triplet are favoured by both $S$ and $U$ parameters. In this region of the parameter space, the 2HDM contribution has an opposite sign compare to the SM one. For sufficiently small pseudoscalar masses (and large triplet masses), this extra contribution could even partially compensate a large positive contribution to the $S$ parameter induced by a heavy ($\approx 300$ GeV) SM Higgs. \textit{A contrario}, scenarios with a very heavy pseudoscalar and a light triplet are disfavoured or even excluded (depending on the actual value of the $T$ parameter) by the experimental upper bounds on $S$.

\subsubsection{Rare $B$ mesons decays}
The $b\to s \gamma$ branching rate measurement \cite{Barberio:2007cr} is known for its stringent bounds on charged Higgs boson masses. The leading contribution of a charged Higgs boson to this loop-induced decay has been derived in Refs. \cite{Grinstein:1987pu,Hou:1987kf}. As expected, the small $\tan\beta$ region for the type I 2HDM \eqref{eq:yuklag2HDMI} is left unconstrained since, in this case, the charged Higgs bosons decouple from the fermions. For larger values of $\tan\beta$, only a very narrow region of the parameter space survives the constraint. In type II 2HDM \eqref{eq:yuklag2HDMII}, the leading order prediction for the lower bound on the charged Higgs mass ($\gtrsim 500$ GeV at 95\% CL) is essentially independent of $\tan\beta$, for $\tan\beta>2$, due to the cancellation between the $\tan\beta$ and the $\cot\beta$ contributions.

However, this leading order prediction of the $b \to s \gamma$ branching ratio suffers from large uncertainties which can be partially reduced by taking into account higher order QCD corrections. For the 2HDM, an estimation of the NLO corrections in \cite{Borzumati:1998tg} shows a sizeable effect which is sufficient to drastically reduce the lower bound obtained at LO for the $H^\pm$ mass. The current status for a type II model is summarised in Ref.~\cite{Misiak:2006zs}. The 95\% (99\%) lower bound on $m_{H^\pm}$ is around 295 (230) GeV and stays practically constant down to $\tan\beta\simeq 2$. Experimental results may even be interpreted as favouring a charged Higgs mass around 650 GeV. As already mentioned, type I scenarios at low $\tan\beta$ values are close to the decoupling limit, such that the new physics corrections are generally small in magnitude (but of opposite sign compared to type II) and the NLO effects are not relevant (e.g.,  see Ref.~\cite{Xiao:2003ya} for a discussion). At higher $\tan\beta$ values, the strong coupling regime is quickly reached and even the NLO prediction may not be well-behaved (\ie , very scale dependent) in this region. In the following, we only consider the $\tan\beta\lesssim 0.5$ region and discard the possibility of a high $\tan\beta$ value.

If the charged Higgs boson couples strongly enough to the $\overline{b}c$ or $\overline{b}u$ quark currents, and at the same time to the $\tau\overline{\nu_\tau}$ leptonic current (\ie , in type II scenarios with large $\tan\beta$), the $B\to D\tau\nu_\tau$ and $B\to \tau\nu_\tau$ branching ratios could be affected. The normalised branching ratio
\begin{equation}
R^{\mathrm{exp}}\equiv \frac{\mathrm{BR}(B\to D\tau\nu_\tau)}{\mathrm{BR}(B\to D l\nu_l)}=(41.6\pm 11.7\pm 5.2)\%
\end{equation}
measured by the BaBar collaboration \cite{:2007ds} sets a 95\% C.L. bound on the type II 2HDM parameter space, namely \cite{Nierste:2008qe}
\begin{equation}
\tan\beta\lesssim 0.23\,\mathrm{GeV}^{-1}\times m_{H^\pm}\,.
\end{equation}
Taking into account the lower bound on $m_{H^\pm}$ from the $b\to s\gamma$ process, this last constraint can only be relevant in the very high $\tan\beta$ region ($\tan\beta\gtrsim 70$) which should anyway be discarded by the requirement of perturbativity for the Yukawa couplings.

A similar bound can be obtained from the rare process $B\to \tau\nu_\tau$. The experimental result from Belle \cite{Ikado:2006un} can be compared to the best SM prediction as follows
\begin{equation}
\frac{\mathrm{BR}(B\to \tau\nu_\tau)}{\mathrm{BR}(B\to \tau\nu_\tau)_{SM}}=1.13\pm 0.44
\end{equation}
where only the experimental error is considered. A normalization with respect to the 2HDM type II model (for a review with Minimal Flavour Violation, see \cite{smith:2008}) gives
\begin{equation}
\tan\beta\lesssim 0.13\,\mathrm{GeV}^{-1}\times m_{H^\pm}\,.
\end{equation}
Assuming $m_{H^\pm}\gtrsim 300$, this constraint may significantly restrict the $\tan\beta\gtrsim 40$ region. Let us however emphasise that this bound should be considered with some caution: the $B\to \tau\nu_\tau$ signal ``evidence'' in the BaBar experiment \cite{:2007xj} is still statistically lower than that obtained by the Belle collaboration, and the theoretical uncertainties associated with the lattice estimate of $f_B$ could be underestimated.

\subsubsection{The $B_0-\overline{B_0}$ mixing}
\begin{figure}[htbp]
\begin{center}
\subfigure[Type I]{\includegraphics[width=0.45\textwidth]{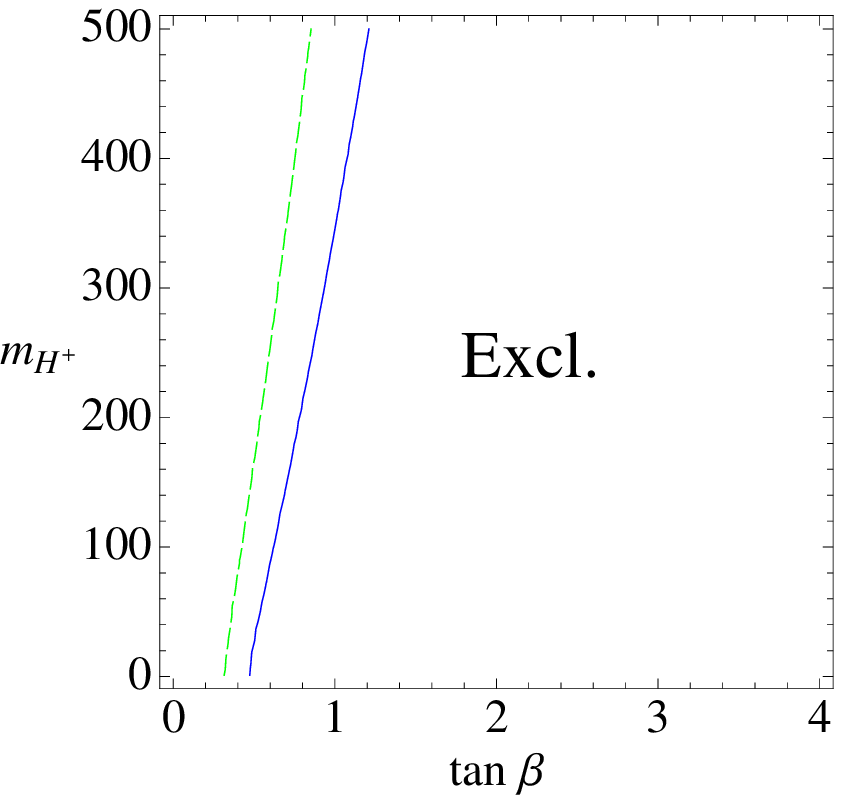}}
\subfigure[Type II]{\includegraphics[width=0.45\textwidth]{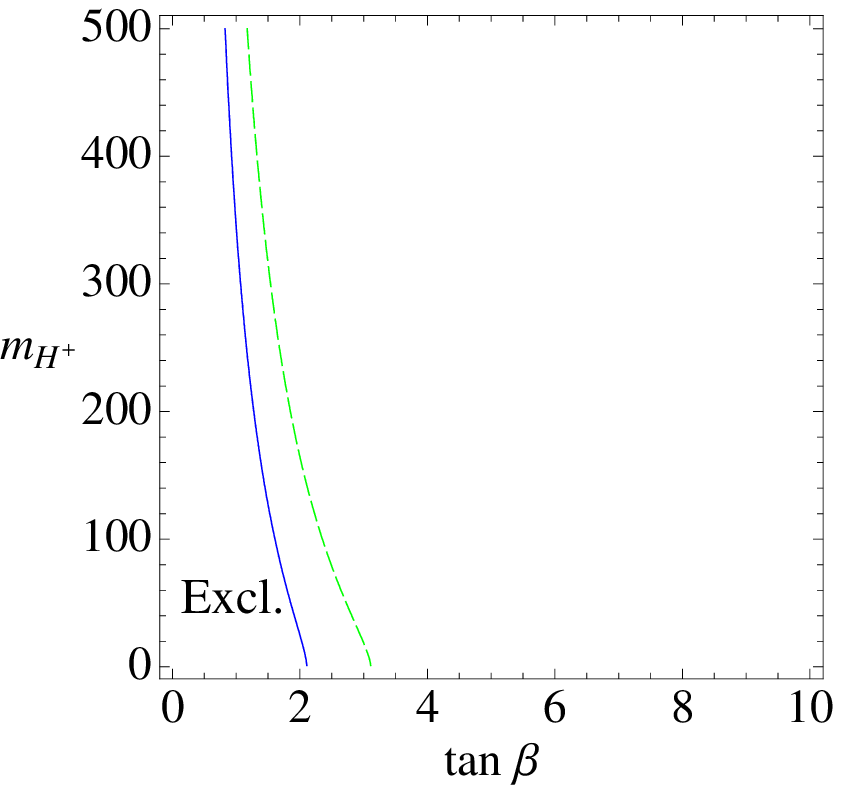}}
\caption{$B_0-\overline{B_0}$ mixing bounds on the charged Higgs mass (in GeV) with respect to $\tan\beta$ in the 2HDM with (a) type I and (b) type II scenarios, at two (dashed green) and three (solid blue) standard deviations. There are no $1\sigma$ limits on these figures due to the slight discrepancy between the SM prediction and the current experimental measurement. Predictions are at leading order.}
\label{fig:b0b0}
\end{center}
\end{figure}

The virtual effects of the charged Higgs bosons on the $B_d-\overline{B_d}$ oscillations have been described at leading order in Ref.~\cite{Geng:1988bq}. The resulting bounds on the charged Higgs mass, with respect to $\tan\beta$, are visible on Fig.~\ref{fig:b0b0}.
Similarly to $b\to s\gamma$, and in order to avoid discussing the actual choice of the SM parameters values entering the $B_0-\overline{B_0}$ mixing, we normalise it to recover the most recent SM prediction from lattice simulations \cite{Ball:2006xx}
\begin{equation}
\Delta m_B^{SM}=0.69\pm 0.15\,\mathrm{ps}^{-1}
\label{eq:deltamBSM}
\end{equation}
to be compared to the experimental world average \cite{Yao:2006px}
\begin{equation}
\Delta m_B^{exp}=0.507\pm 0.005\,\mathrm{ps}^{-1}\,.
\label{eq:deltamBexp}
\end{equation}
Consequently, the error on the resulting constraint is clearly dominated by the theoretical uncertainty in \eqref{eq:deltamBSM}. For type I scenarios, the $\tan\beta\lesssim 0.5-1$ constraint obtained on the whole range for the charged Higgs mass is similar to the one obtained from the $b\to s\gamma$ process. In the case of type II 2HDM, the $\tan\beta<2$ region is excluded at more than 95\% CL, almost independently of $m_{H^\pm}$. Like in the $b\to s\gamma$ case, the inclusion of $\mathcal{O}(\alpha_s)$ QCD corrections reduces the sensitivity of $\Delta m_B$ to charged Higgs contributions and slightly weakens the above constraints \cite{Urban:1997gw}.

\subsubsection{The $Zb\overline{b}$ vertex}
Loop corrections involving new charged and neutral scalars may give sizable contributions to $R_b$, the observable hadronic branching ratio of $Z$ bosons to $b\overline{b}$, and to $A_b$, the $b$-quark asymmetry. These corrections have been derived in great detail in Ref.~\cite{Logan:1999if} for the $CP$ conserving 2HDM. Corrections to $R_b$ and $A_b$ are expected to have similar magnitudes. However, the high experimental precision associated with the $R_b$ measurement makes it much more discriminating than $A_b$ on the whole parameter space, as shown by an explicit numerical analysis. 

\begin{figure}[htbp]
\begin{center}
\subfigure[$\tan\beta=30$]{\includegraphics[width=0.45\textwidth]{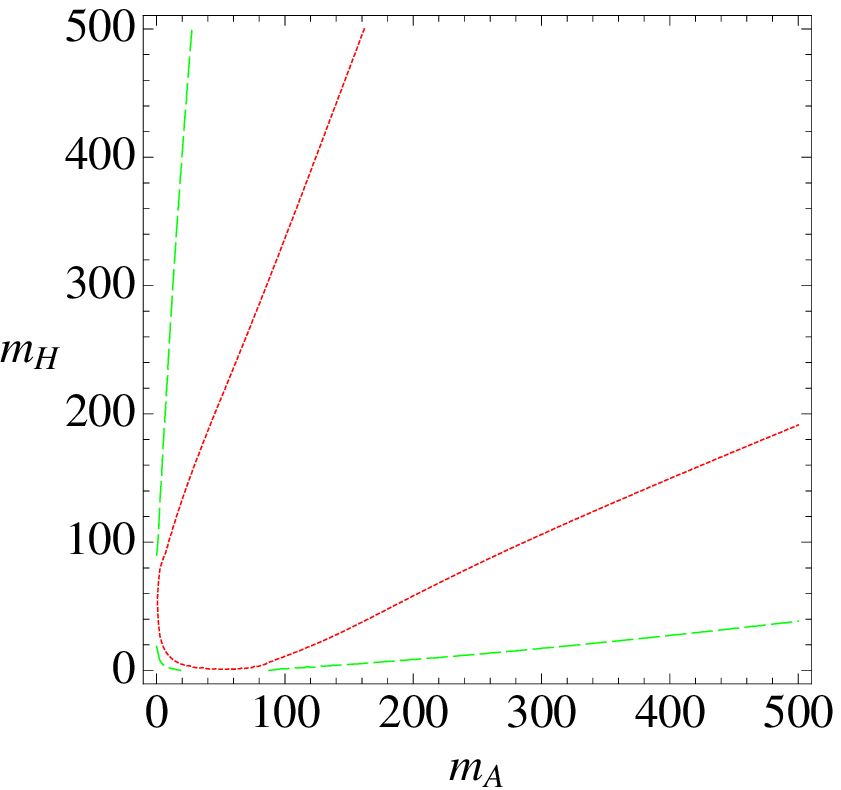}}
\subfigure[$\tan\beta=50$]{\includegraphics[width=0.45\textwidth]{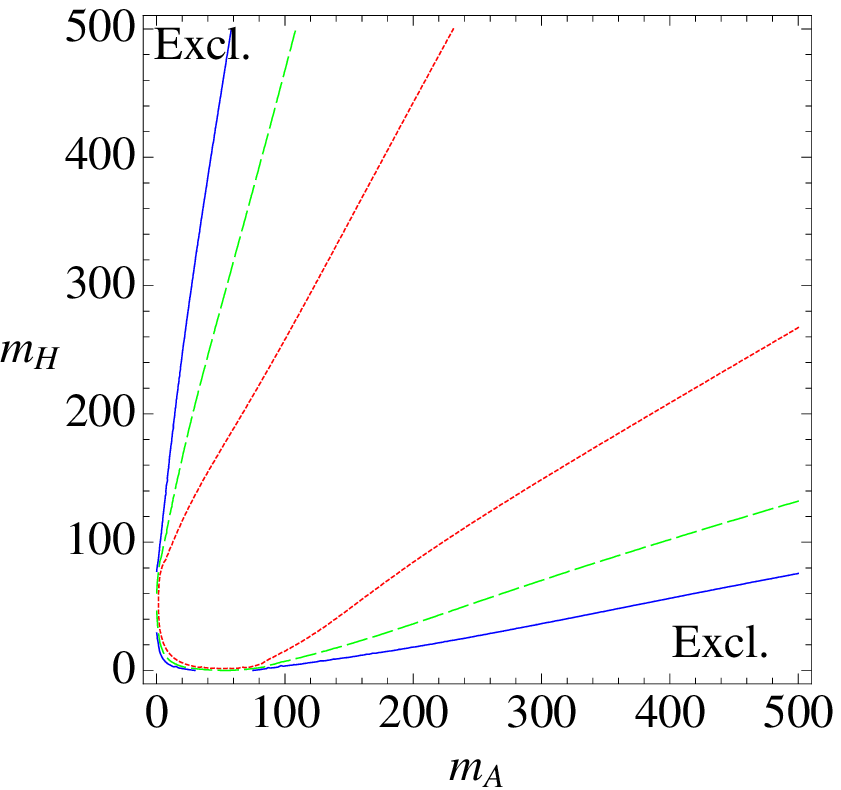}}
\caption{Constraints at $1\sigma$ (dotted red), $2\sigma$ (dashed green) and $3\sigma$ (solid blue) on the type II M2HDM parameter space from the $R_b$ measurement, for two different values of $\tan\beta$.}
\label{fig:Rb}
\end{center}
\end{figure}

In type I models, the only relevant contribution is the charged Higgs correction to $g_L$ since all the other ones are suppressed by $m_b/v$. However, the typical bound extracted in this case, namely $\tan\beta\lesssim 1$, is not competitive with the one coming from the $B_0-\overline{B_0}$ measurement. The situation is more interesting in type II models where the neutral scalar contributions are potentially sizable. In the type II M2HDM, this restricts the allowed region in the $(m_{A^0},m_{H^0})$ plane, as illustrated on Fig.~\ref{fig:Rb}. 
Since the $(H^0,H^\pm)$ triplet is forced to be rather heavy by the $B$ physics constraints previously reviewed, the mass of the pseudoscalar $A^0$ is bounded from below. For $\tan\beta=50$ and $m_{H^\pm}>300$ GeV, for example, the bound is approximatively $m_{A^0}>60$ GeV at 95\% CL. This bound is somewhat lower than the one obtained in Ref.~\cite{Logan:1999if} ($m_{A^0}>100$ GeV) in the days when the experimental value of $R_b$ was sizably larger, leaving less room for negative contributions from new physics.

\subsubsection{The muon anomalous magnetic moment}
\label{subsec:gminus2}
The anomalous magnetic moment of the muon is known to be a rather interesting probe for BSM physics. The SM theoretical prediction, incorporating the $e^+e^-\to\pi\pi$ data obtained by CMD-2, KLOE and SND, gives \cite{Hagiwara:2006jt}
\begin{equation}
a_\mu^{\mathrm{th,SM}}=(11659180.4\pm 5.1)\times 10^{-10}
\end{equation}
to be compared to the experimental measurement from the Brookhaven experiment \cite{Bennett:2006fi}
\begin{equation}
a_\mu^{\mathrm{exp}}=(11659208.0\pm 6.3)\times 10^{-10}\,.
\end{equation}
The $3.4\sigma$ deviation between these two values may be optimistically interpreted as a signal of new physics, or, at least, as a valuable constraint. Due to the smallness of the muon mass, this constraint is only relevant for Higgs physics when the coupling to the leptons is increased compared to its SM value. In the M2HDM considered here, this corresponds to the type II scenario for Yukawa couplings, for which the one- and two-loop contributions are reported in Ref.~\cite{Krawczyk:2001pe}.

\begin{figure}[htbp]
\begin{center}
\subfigure[]{\includegraphics[width=0.45\textwidth]{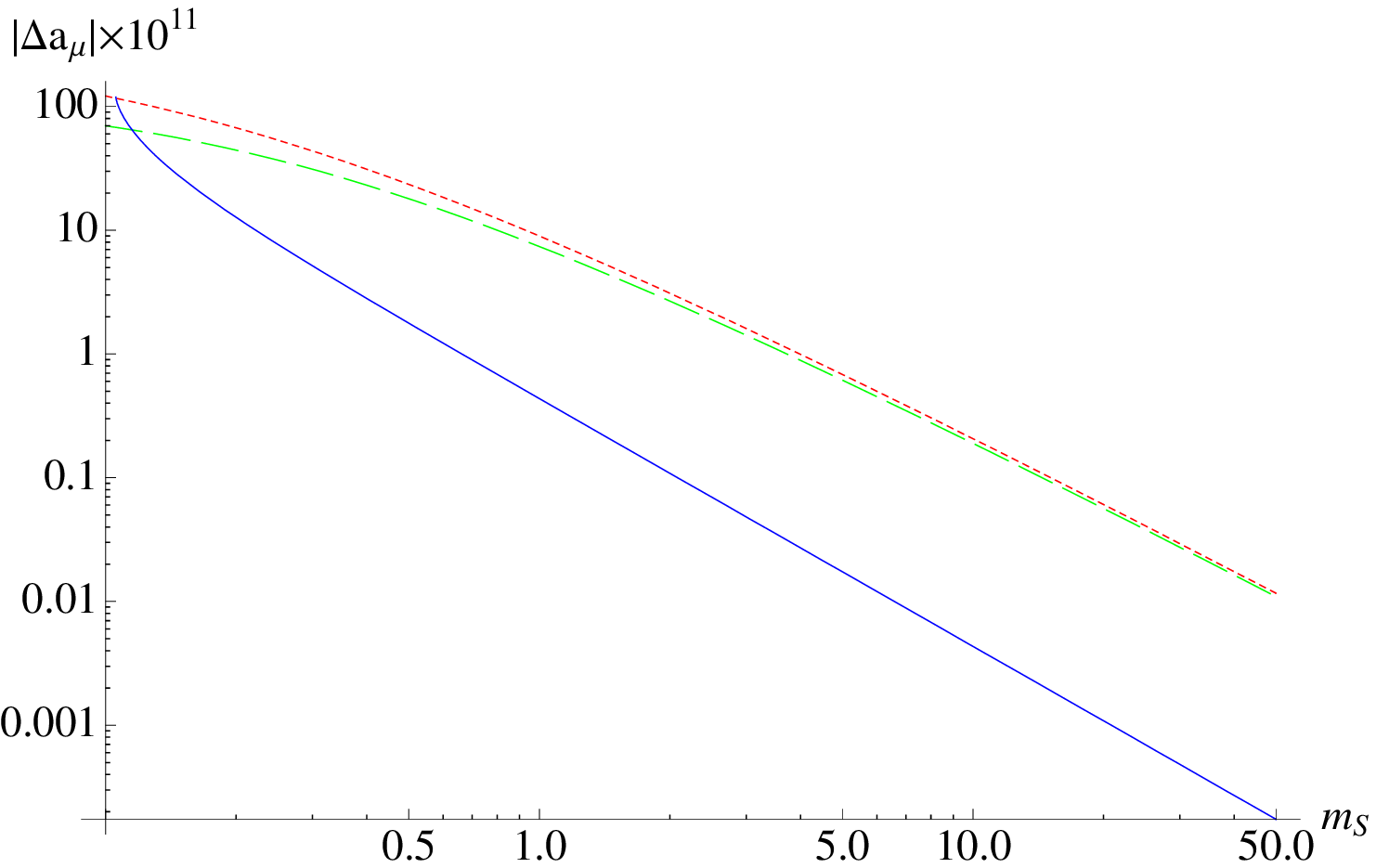}}
\subfigure[]{\includegraphics[width=0.45\textwidth]{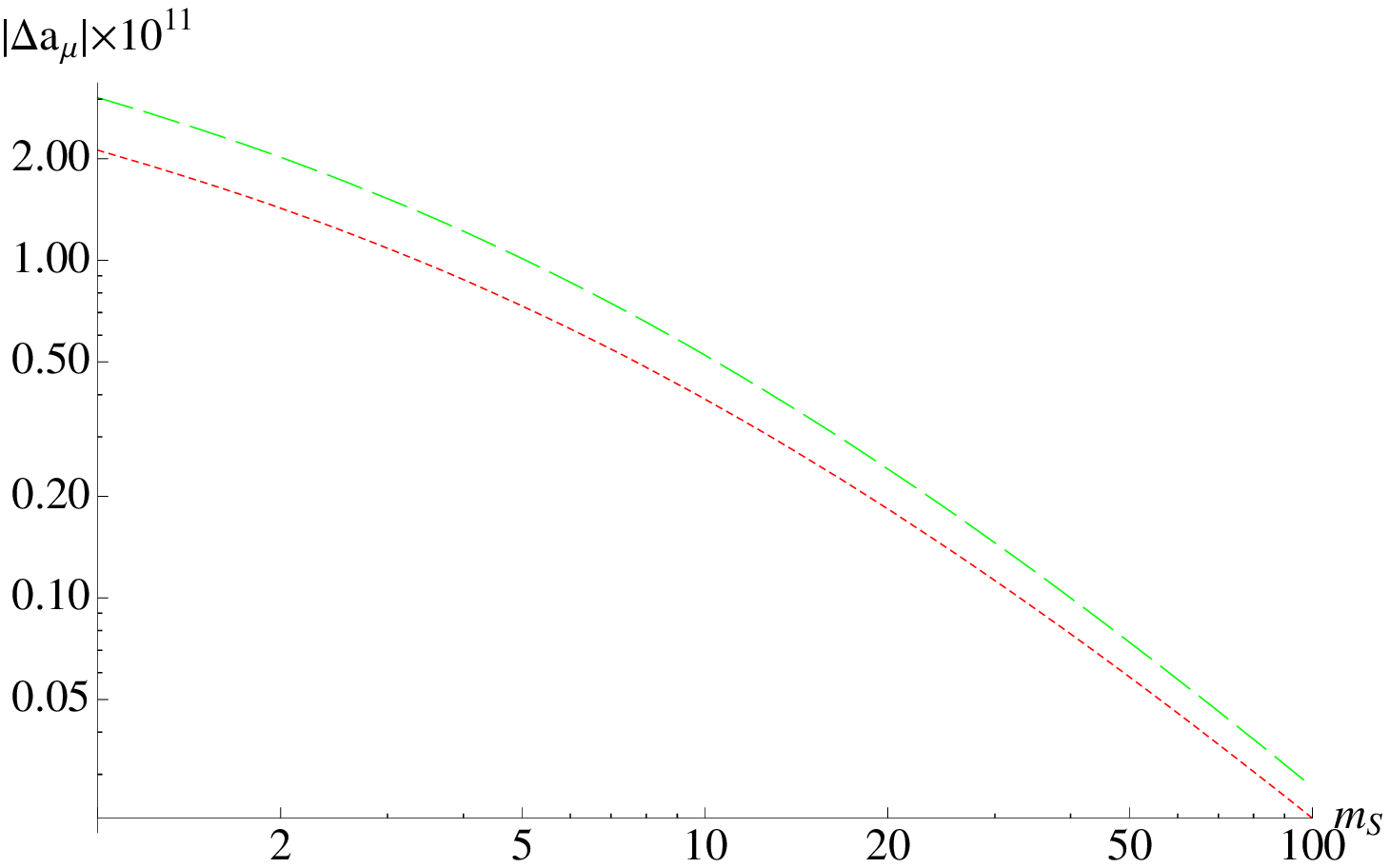}}
\caption{(a) Absolute value of the one-loop contribution to $a_\mu$ from a neutral scalar $H^0$ (dotted red), a pseudoscalar $A^0$ (dashed green) and a charged Higgs boson (solid blue), if $\tan\beta=1$. The $H^0$ contribution is positive while the $A^0,H^\pm$ are negative.  (b) Same for the two-loop contributions from a neutral scalar $H^0$ (dotted red) and a pseudoscalar $A^0$ (dashed green). At two-loop, the neutral scalar contribution is negative while the pseudoscalar one is positive. Only the $b$, $\tau$ and $\mu$ fermion loops are included.}
\label{fig:amu}
\end{center}
\end{figure}

At one-loop, the scalar contribution $a_\mu^{H^0}$ is positive whereas the pseudoscalar and the charged Higgs boson give negative contributions. Each contribution reaches its extremum at small masses and vanishes like $m_\mu^2/m_S^2\log(m_S^2/m_\mu^2)$ at large masses. The absolute magnitude of each type of contribution is shown on Fig.~\ref{fig:amu}(a) for $\tan\beta=1$.
The total one-loop correction is dominated by the contributions of neutral Higgses for masses above 0.2 GeV. Solving at the one-loop level the $a_\mu$ theory/experiment discrepancy in the M2HDM with a moderate $\tan\beta$ would require a very light ($\lesssim 10$ GeV) \textit{scalar} $H^0$. The situation changes when considering the dominant two-loop contributions. The $H^0$ contribution is now negative whereas the $A^0$ one is positive. The  total two-loop contributions can be seen in Fig.~\ref{fig:amu}(b). By comparing with Fig.~\ref{fig:amu}(a), one finds that these two-loops corrections are dominant for $m_S\gtrsim 10$ GeV (they cancel against the one-loop part for $m_S\approx 5$ GeV). Due to the opposite sign of the scalar and pseudoscalar contributions, solving the $a_\mu$ theory/experiment discrepancy in the M2HDM with a moderate $\tan\beta$ in this region would require a light \textit{pseudoscalar} ($20\lesssim m_{A^0}\lesssim 100$ GeV). If $\tan\beta\approx 30$, a very good agreement can even be reached for $m_{A^0}\approx 20$ GeV. A larger $\tan\beta$ value would require a heavier pseudoscalar and \textit{vice versa}. In all cases, corrections to $a_\mu$ clearly differ when considering either scalar or pseudoscalar particles. The $a_\mu$ measurement can thus potentially help to physically disentangle the usual and twisted custodial scenarios introduced in Section~\ref{sec:model}.

\subsection{Collider constraints}
\subsubsection{LEP}
LEP searches for a SM Higgs boson in the standard decay modes $h^0\to b\overline{b}$ and $h^0\to\tau^+\tau^-$, have provided an exclusion at a 95\% confidence level for a Standard Model Higgs boson with a mass lower than 114.4 GeV \cite{Barate:2003sz}. In the context of models with an extended scalar sector, $h^0$ may also decay, possibly dominantly, in a pair of lighter Higgs bosons. Since such a light object has escaped the LEP searches, it must either have a reduced coupling to $ZZ$ or unusual decay properties. In the context of the M2HDM, both $A^0$ and $H^0$ could satisfy this requirement thanks to their vanishing couplings to $Z$ boson pairs. However, here $H^0$ is nearly degenerate in mass with $H^\pm$. Since the possibility of a light ($\lesssim$ 100 GeV) charged Higgs boson is strongly disfavoured by both direct and indirect measurements, we then focus on the light $A^0$ hypothesis in the following.

In the forthcoming phenomenological analyses, we conservatively consider the usual SM bound $m_{h^0}>114.4$ GeV. However, it should be mentioned that this bound could be slightly lowered if the $h^0\to b\overline{b}$ branching ratio was reduced due to the presence of the $h^0\to A^0 A^0$ decay mode. Constraints on this decay mode from LEP data have been considered, for example, in the framework of the Next-to-Minimal Supersymmetric Standard Model (see \cite{Accomando:2006ga} and reference therein). If $m_{A^0}>2 m_b$, it has been first thought that this decay could explain the simultaneous excesses observed in the $Z2b$ \cite{Barate:2003sz} and $Z4b$ \cite{Schael:2006cr} final states by adjusting $m_{h^0}$, $\mathrm{BR}(h^0\to A^0 A^0)$, $\mathrm{BR}(h^0\to b\overline{b})$ and $\mathrm{BR}(A^0\to b\overline{b})$. However, the $Z4b$ excess tends to favour slightly higher masses for $h^0$ (in the 105-110 GeV region) compared to the main $Z2b$ excess (in the 100 GeV region), decreasing the significance of a global fit.

The model independent searches for the $e^+e^-\to Z^*\to H^0 A^0$ pair production process at LEP (in the $4b$, $2b2\tau$ and $4\tau$ channels) put the tightest constraint on the M2HDM mass spectrum, and in particular on the type I model where indirect constraints from $B$ physics still allow for a relatively light triplet $(H^\pm,H^0)$. Taking into account the values of $\mathrm{BR}(H^0,A^0\to b\overline{b})$ and $\mathrm{BR}(H^0,A^0\to \tau^+\tau^-)$, the final result from \cite{Schael:2006cr} can be roughly summarized as the bound $m_{A^0}+m_{H^0}\gtrsim 170$ GeV for $m_{A^0}\approx m_{H^0}$. If $A^0$ is very light, namely $m_{A^0}\lesssim 30$ GeV, a slight loss of efficiency in the $4b$ and $2b2\tau$ analysis allows for $m_{H^0}\gtrsim 130$ GeV. 

If $m_{H^0}\gtrsim m_Z+m_{A^0}$, the scalar $H^0$ could also decay into $Z A^0$, thus reducing dramatically the branching ratio of $H^0\to \tau^+\tau^-,b\overline{b}$. The final state signatures associated with this possibility (namely $Z4b$, $Z2b2\tau$ and $Z4\tau$, see also Ref.~\cite{Dermisek:2008id} for a discussion in the small $\tan\beta$ MSSM) would mimic the $h^0\to A^0 A^0$ process described previously, but with a  different kinematical structure.
In this case, the decay products of the $Z$ boson and the two \textit{softer} $b$-jets should be used to reconstruct the $H^0$ mass. In the absence of any dedicated experimental study for this open possibility, we adopt a conservative approach and restrict the model using the $(m_{H^0},m_{A^0})$ limits already mentioned.

Another strong constraint in the low $m_{A^0}$ region for type II models can be deduced from searches for the Yukawa-induced process $e^+e^-\to b\overline{b} A^0$ with $A^0\to \tau^+\tau^-,b\overline{b}$. The result for each mode is available in Ref.~\cite{Abdallah:2004wy}. Due to the reduced branching ratio for $A^0\to\tau^+\tau^-$, the $A^0\to b\overline{b}$ mode is the most restrictive one in the $m_{A^0}>2m_b$ region. The limit is $\tan\beta\lesssim 20$ for $m_{A^0}\simeq 10$ GeV and quickly becomes less relevant for higher masses due to the smaller production cross section. The loop decay $Z\to A^0 \gamma$ (through a quark or a lepton loop) could in principle also be used to constrain the $A^0$ mass. However, an extensive analysis of this channel \cite{Krawczyk:1998kk} has shown that the LEP measurement sensitivity (of order $10^{-6}$ for the associated branching ratio) was not sufficient to put a lower bound on $m_{A^0}$ tighter that the one obtained from the Yukawa-induced process.

Finally, results for  charged Higgs boson searches at LEP in the general 2HDM (type I and type II) by LEP collaborations \cite{Abdallah:2003wd,:2008be}, motivated by the possibility for three-body decays of Higgs bosons reviewed in \cite{Akeroyd:1998dt}, are also considered.
In addition to the usual fermionic decays $H^+\to\tau^+\nu_\tau$ and $H^+\to c\overline{s}$, the possibility for charged Higgs boson pair production ($e^+e^-\to Z^*\to H^+H^-$) with the decay into $W^{+(*)}A^0$ has also been taken into account. The existence of a charged Higgs boson with mass lower than 76.7 GeV (type I) or 74.4 GeV (type II) is excluded at the 95\% CL, for a wide domain of the parameter space. 

\subsubsection{Tevatron}
To date, the total integrated luminosity collected at the Tevatron collider is not sufficient yet to exclude a Standard Model Higgs boson at masses higher than the LEPII limit, except for a narrow window around 165 GeV \cite{Tev2009Higgs}. We thus focus here on the $h^0\to A^0 A^0$ exotic decay, with $A^0\to b\overline{b},\tau^+\tau^-$. 

The case of direct production of $h^0$ through its effective coupling to gluons, followed by the decay $h^0\to A^0 A^0\to 4b$ has been covered in Ref.~\cite{Stelzer:2006sp}. Unfortunately, the $4b$ QCD background overwhelms the signal and a discovery could only be achieved if the $h^0$ production would be enhanced by at least one order of magnitude. The same process where one of the $A^0$ decays into $\tau^+\tau^-$ instead of $b\overline{b}$ could provide a cleaner signature, at the price of reducing the rate due to the small $A^0\to \tau^+\tau^-$ branching ratio.

Since the associated production of $h^0$ with a vector boson $V=W,Z$ is the second largest production mechanism at the Tevatron, it is also natural to consider the exotic decay $h^0\to A^0 A^0$ in this context. Detailed studies (see Refs. \cite{Cheung:2007sva}, \cite{Djouadi:2008uw}, \cite{Carena:2007jk} and the references therein for an overview) of both the $V2b2\tau$ and $V4b$ final states have proven the potential for such signature. However, the associated statistical significance is too small to constrain the model parameters with the currently available integrated luminosity.

Searches for the $H^0$ and $A^0$ bosons at the Tevatron experiment take place for production in association with $b$ quarks, or in gluon fusion, and decays to $b\overline{b}$ and $\tau^+\tau^-$ final states. Since most analyses are oriented towards the MSSM Higgs bosons discovery, they focus mainly on the $m_S\gtrsim 70$ GeV mass region. Exclusion regions for Run I \cite{Roco:2001} and Run II \cite{CDF:2007} can be (very) conservatively translated as a constant $\tan\beta<35$ bound on the whole $ m_{A^0,H^0}> 70$ GeV mass range.

Another interesting possibility to produce a light $A^0$ boson at the Tevatron (and possibly at the LHC) is the charged Higgs associated production $p\overline{p}\to W^{\pm*} \to H^\pm A^0$, where the charged Higgs may further decay into $W^\pm A^0$, which has been proposed in Ref.~\cite{Akeroyd:2003jp} and discussed more recently in Ref.~\cite{Akeroyd:2007yj} in the context of the NMSSM.
In the most favourable mass scenarios ($m_{A^0}<20$ GeV, $m_{H^\pm}<90$ GeV), the associated cross sections can be larger than 500 fb and a signal could be detected in the $W4b$ final state. 

Finally, if $m_{H^\pm}< m_t$, a top quark could have a sizable branching ratio to $H^+b$. This possibility has been considered by the Tevatron experiments for various charged Higgs decay hypothesis. The result is a 95\% CL upper bound on the $t\to H^+b$ branching ratio,{\it  i.e.}, $\lesssim 0.2$ for the whole $H^+$ mass range \cite{CDF2008Charged}.

\subsection{Constraint summary and mass spectrum}
All the relevant constraints introduced in the previous sections are summarised in Tab.\ref{tab:constr}. 
\begin{table}
\begin{center}
\begin{tabular}{|l|c|l|}
  \hline
  \multirow{5}{*}{Type I \& II} &  $m_{h^0}<500$ GeV & Unitarity\\
   \cline{2-3}
   & $m_T^2-m_A^2<(400\,\mathrm{GeV})^2<m_{h^0}^2$ & Perturbativity\\
   \cline{2-3}
   & $m_{H^\pm}=m_{H^0}+\epsilon$ & $\Delta T\approx 0$, if $m_{h^0}>250$ GeV \\
   \cline{2-3}
   & $m_{A^0}\ll m_T$ & $\Delta S\approx 0$, if $m_{h^0}>250$ GeV\\
   \cline{2-3}
   & $m_{h^0}>114.4$ GeV & LEP bound on the SM Higgs\\
  \hline\hline
  \multirow{4}{*}{Type I} & $m_T\gtrsim 130$ GeV & LEP $Z\to H^0 A^0$ ($m_{A^0}< 30$ GeV)\\
  \cline{2-3}
  & $m_T\gtrsim 170\,\mathrm{GeV}-m_{A^0}$ & LEP $Z\to H^0 A^0$ ($m_{A^0}> 30$ GeV)\\
  \cline{2-3}
  & $m_{A^0}> 10$ GeV & No fine tuning in \eqref{eq:A0H0massTC}\\
  \cline{2-3}
  & $\tan\beta\lesssim 0.4$ & $b\to s\gamma$ and $B_0-\overline{B_0}$ mixing\\
  \hline\hline
  \multirow{5}{*}{Type II} & $m_T>295$ GeV & $b\to s\gamma$ \\
  \cline{2-3}
  & $m_{A^0}\gtrsim 30$ GeV & $R_b$ and LEP $bbA\to 4b$\\
  \cline{2-3}
  & $m_{A^0}\lesssim 100$ GeV & Favoured by $a_\mu$\\
  \cline{2-3}
  & $\tan\beta\gtrsim 5$ & $B_0-\overline{B_0}$ mixing \\
  \cline{2-3}
  & $\tan\beta\lesssim 35$ & $B\to\tau \nu_\tau$ and LEP $bbA\to 4b$ \\
  \hline
\end{tabular}
\end{center}
\caption[]{Summary of the relevant constraints for the M2HDM defined by Eqs.~\eqref{eq:twistedZ2pot}, \eqref{eq:yuklag2HDMI} and \eqref{eq:yuklag2HDMII}. \label{tab:constr} }
\end{table}
Only the most stringent bounds are presented, and some of them could be strongly correlated with others. These constraints have been derived with the implicit assumption that all BSM contributions beyond the M2HDM are negligible. They thus appear more as guidelines to restrict the parameter space for the phenomenological study of collider signatures presented in Chapter 4, than as strict bounds. 

One can foresee qualitative differences between the ``twisted'' scenario and other 2HDM realisations. The (constrained) $CP$-conserving MSSM Higgs phenomenology in the low mass region is very similar to what can be expected in the SM. So, an identification of its enlarged scalar sector at the LHC will have to rely on the direct detection of the heavy states $H^\pm$, $H^0$ and $A^0$, which may require a high luminosity ($\sim 100$ fb$^{-1}$, depending on $\tan\beta$). The same conclusion also holds for most of the ``usual'' scenarios considered in the literature, where the nearly degenerate custodial triplet $(H^\pm,A^0)$ is forced to be relatively heavy by severe $B$ meson physics constraints.

The situation could be completely different in the M2HDM considered here, due to its distinctive inverted spectrum (see Eq.~\eqref{eq:invspec}). In type I scenarios, the reduced Yukawa coupling of the charged Higgs allows for a moderately light custodial triplet $(H^\pm,H^0)$. A small mass splitting inside this triplet (and, to a lesser extent, the presence of a light pseudoscalar $A^0$) is required to help a rather heavy SM Higgs (say, $m_{h^0}\approx 300$ GeV) in passing the electroweak precision tests. If $m_{h^0}>2m_T$, many exotic scalar decays can be kinematically allowed ($h^0\to H^+ H^-$, $h^0\to H^0 H^0$, \ldots) giving rise to unconventional collider signatures.
In type II scenarios, the custodial triplet mass is constrained to be larger due to the $b\to s\gamma$ bound on the charged Higgs mass. A relatively small mass splitting inside this heavy triplet (of order 5\%) is enough to allow for a very heavy SM Higgs\footnote{Even though it appears as an interesting possibility required in the context of the M2HDM to satisfy Eq.~\eqref{eq:hierarchyscal}, let us emphasize the presence of a heavy SM Higgs boson plays no role in the phenomenological signatures of type II scenarios considered in Sections \ref{bbza} and \ref{chargedhiggs}.}($m_{h^0}\approx 400$ GeV). The electroweak oblique parameter $S$ together with the available data for the muon anomalous magnetic moment slightly favour the presence of a light pseudoscalar in this case but its mass is bounded from below ($m_{A^0}>30$ GeV) by the LEP direct searches and the $R_b$ measurement. Even though the $m_{h^0}>2m_T$ condition is hardly satisfied in this context due to unitarity and perturbativity constraints on $m_{h^0}$, unusual decays for $H^0$ and $H^\pm$ remain an interesting open possibility.

\section{Discovery potential at the LHC}
\label{sec:LHCpheno}
This last section is dedicated to the discovery potential of type I and II M2HDMs at the Large Hadron Collider (LHC). First, we review the main Higgs production and decay modes for both type I and type II scenarios. Next, the constraints derived in Section~\ref{sec:constr} are used to identify two benchmark points in the allowed regions of the model parameter space, one for each type. Finally, after a short description of simulation and reconstruction tools, the prospects for three particularly relevant signal processes at the LHC are reviewed in more detail.

\subsection{Production and decay modes of Higgs bosons and benchmark points}

\subsubsection{Type I scenarios}
The type I M2HDM is characterised by a $\tan\beta$ parameter smaller than unity. In this case the only relevant direct production mechanisms for $H^0$, $A^0$ and $H^\pm$ involves couplings to top quarks 
and/or gauge interactions (e.g., $q\overline{q}\to Z^* \to H^0 A^0$), while all the usual SM production modes are still relevant for the $h^0$.

In general, the cross sections for the production of pairs of scalar particles barely reach 1 pb. For $H^\pm H^\mp$ and $H^0H^0$, it is even smaller due to the constraint on their mass to be heavier than 130 GeV. 
Around this mass bound, the dominant production mechanism for the $ H^\pm$ is  via the top decay $t\to H^+ b$ due to the very large $t\overline{t}$ production cross section at the LHC,

If $m_{H^\pm}\gtrsim 150$ GeV, the associated production of the charged Higgs boson with a single top quark becomes dominant compared to the $t\to H^+ b$ decay which goes down quickly as $m_{H^\pm}$ gets closer to the kinematical bound $m_t-m_b$. Finally, the production  of neutral Higgs bosons associated with the $t\overline{t}$ pair is also possible but cross sections are smaller than for the SM Higgs boson. 

\begin{figure}[htbp]
\begin{center}
\subfigure[$m_{A^0}=30$ GeV]{\includegraphics[width=0.49\textwidth]{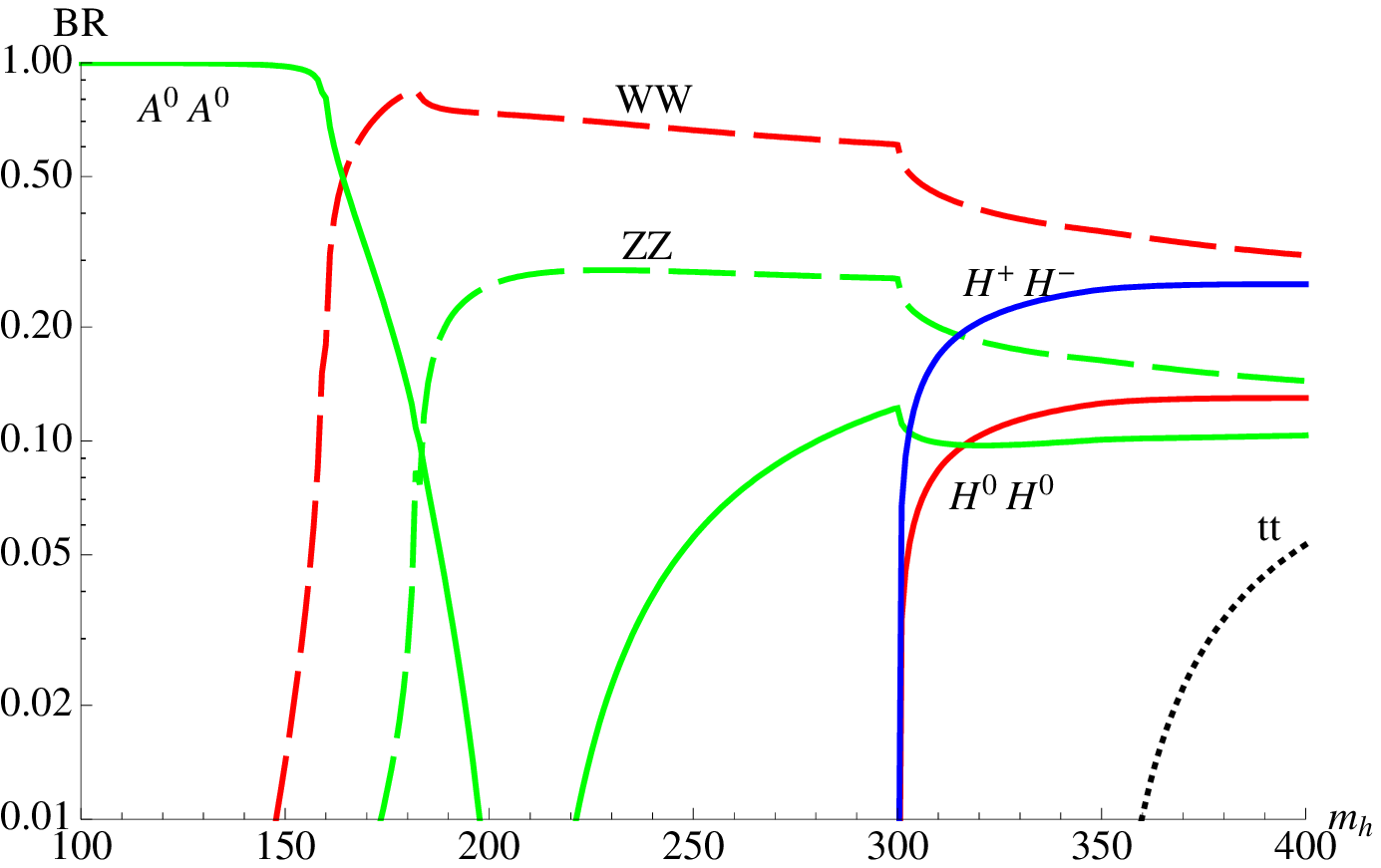}}
\subfigure[$m_{A^0}=70$ GeV]{\includegraphics[width=0.49\textwidth]{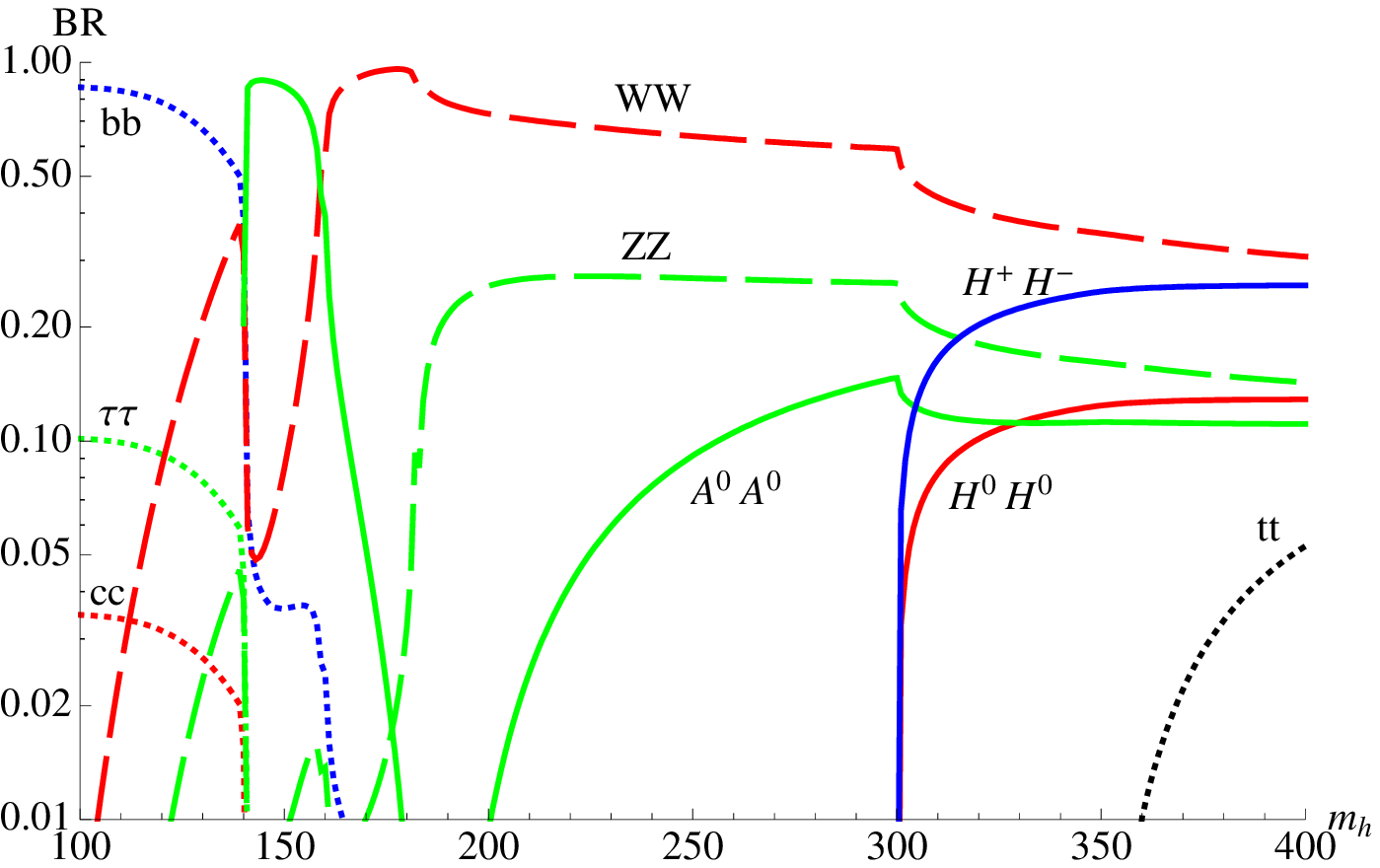}}
\caption{Tree-level branching ratio of $h^0$ into fermions (dotted lines), vector bosons (dashed lines) and scalars (solid lines), with respect to its mass and for two different values of $m_{A^0}$.}
\label{fig:plotbr_1_h0}
\end{center}
\end{figure}

The main decay modes of the $h^0$, shown in Fig.~\ref{fig:plotbr_1_h0}
for two different values of the $A^0$ mass, can be qualitatively very different from those observed in the SM or in the MSSM. In the low mass region,{\it  i.e.}, if $m_{h^0}<2m_{A^0}$, the $h^0$ primarily decays into $b$, $\tau$ and $c$ pairs (in order of importance) like in the SM.  If  $2m_{A^0}< m_{h^0}< 2m_W$, the dominant decay mode is $h^0\to A^0 A^0$ due to the large trilinear scalar coupling. Note however  that this coupling, and hence the $h^0\to A^0 A^0$ branching ratio, vanishes in the narrow mass window around  $m_{h^0}=\sqrt{2(m_T^2-m_{A^0}^2)}$. For $2m_W<m_{h_0}<2m_{T}$, the decay into a pair of gauge bosons dominates with BRs similar to the SM ones. Above the $2m_T$ threshold, both the $h^0\to H^0 H^0$ and the $h^0\to H^+ H^-$ decays are kinematically allowed and account for about half of the total width. This total decay width is sizably larger than in the SM at low masses (around 1 GeV for $m_{h^0}=100$ GeV), due to the unusual $h^0\to A^0 A^0$ decay mode, but it remains of the same order of magnitude at higher masses.

For the other Higgs bosons, $A^0$, $H^0$ and $H^\pm$, the only possible decays are into SM particles (scaled by $\tan\beta$) if $m_{H^\pm/H^0}<m_{W^\pm/Z}+m_{A^0}$. If the $H^0$ ($H^\pm$) mass is above the $ZA^0$ ($W^\pm A^0$) mass threshold, the branching ratio (BR) into these particles become close to unity. For $H^0$ and $H^\pm$, the total decay widths remain at most of the order of the SM Higgs one at the same mass.

\subsubsection{Type II scenarios}
All production mechanisms for $h^0$, and those involving only gauge or scalar interactions for $H^0$, $A^0$ and $H^\pm$, are identical to those described above for type I scenarios. The main difference is the enhanced coupling of the extra Higgs bosons to down-type quarks and to charged leptons. The direct production through $b$ quark annihilation thus becomes by far the main production mechanism at the LHC for the neutral Higgs bosons $H^0$ and $A^0$.

The $h^0$ couplings remain the same in type II and type I models, hence the branching ratio patterns shown on Fig.~\ref{fig:plotbr_1_h0} remain valid. The $H^0$ and $H^\pm$ decays are still dominated by the $ZA^0$ and $W^\pm A^0$ at high masses. However, contrary to what happens for type I models, the decays into $b\overline{b}$ and $tb$ respectively are not negligible and become dominant for masses below 250 GeV. 

\subsubsection{Benchmark points}

\begin{figure}[htbp]
\begin{center}
\subfigure[Type I]{\includegraphics[width=0.49\textwidth]{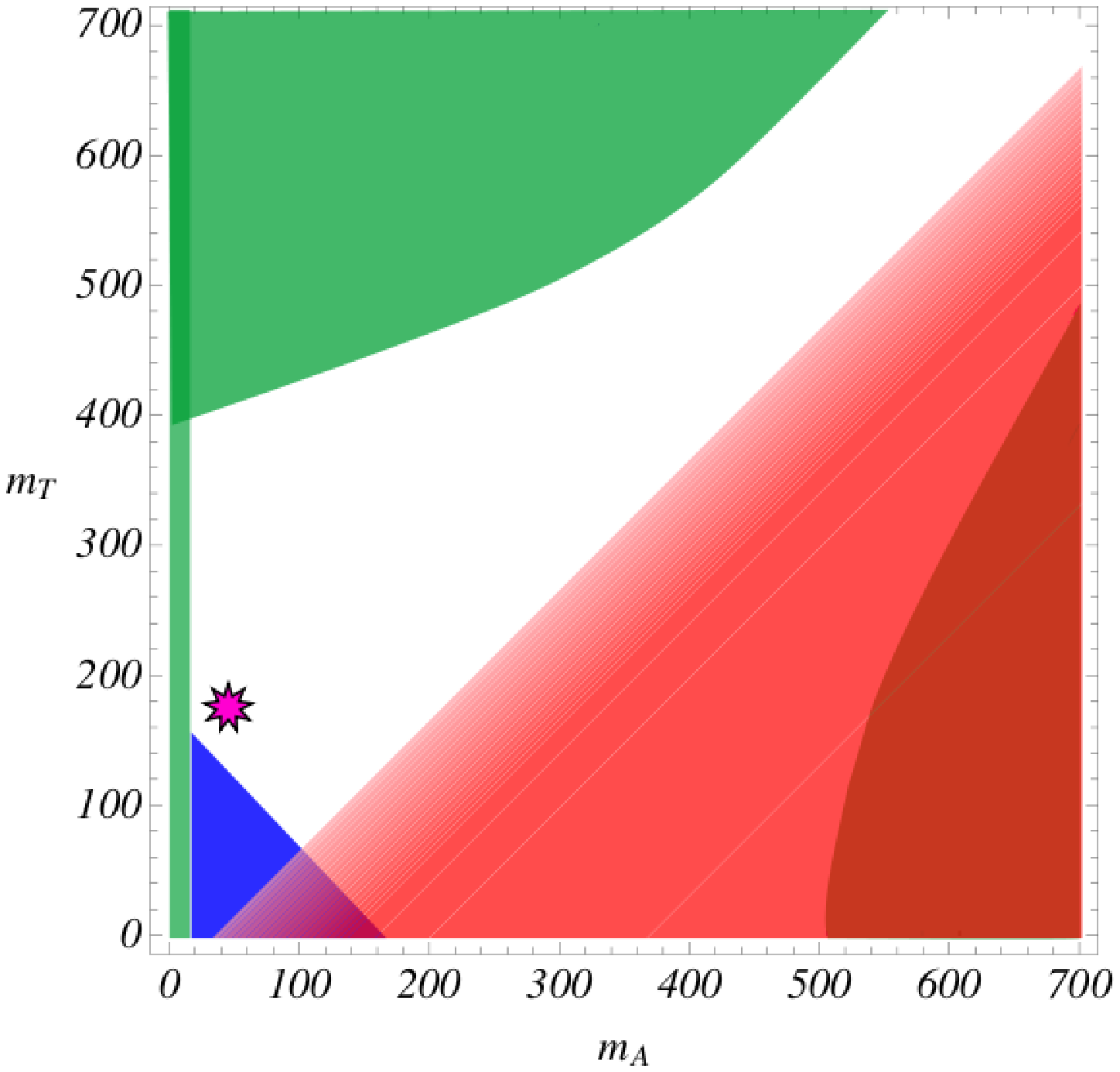}}
\subfigure[Type II]{\includegraphics[width=0.49\textwidth]{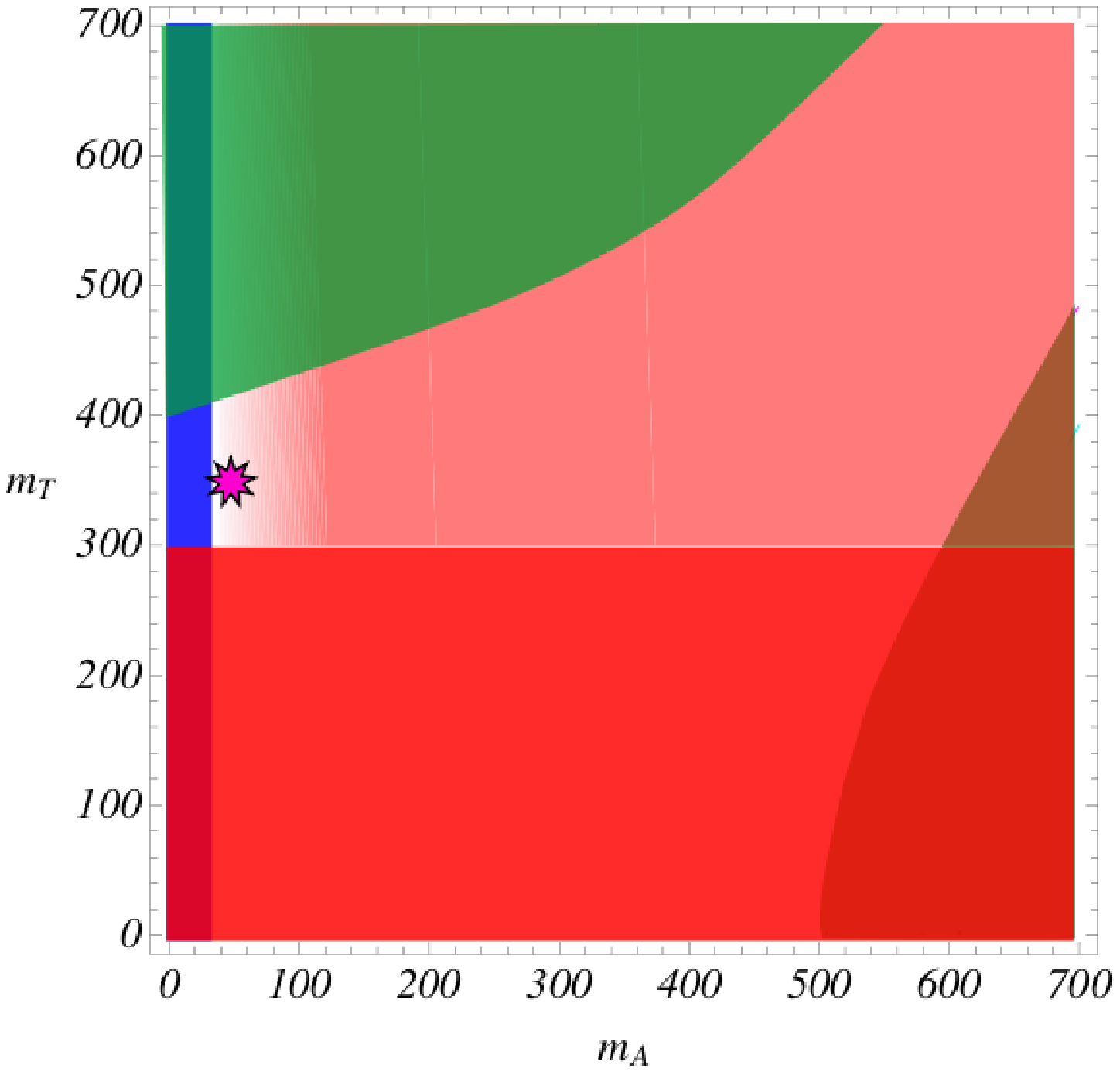}}
\caption{Summary of all relevant theoretical (green), indirect (red) and direct (blue) constraints on the type I and type II M2HDM, in the plane [$m_A$,$m_T$]. Gradient bounds indicate milder indirect constraints, like constraints associated with the $S$ and $a_\mu$ parameters, which should not be considered too strictly. The purple stars indicate the two choices of benchmark points as detailed in the text.}
\label{fig:constr}
\end{center}
\end{figure}

The identification of the unconstrained region for the parameter space of type I and type II scenarios drives the choice of benchmark points in the Higgs inverted mass spectrum (see Eq. \eqref{eq:invspec}). The possible values for $m_T$ and $m_A$, taking into account direct and indirect constraints, are summarized in Fig.~\ref{fig:constr}. 
In this work  we restrict ourselves to two representative cases, one for each type: ``BP1" for the type I and ``BP2" for the type II. For each case, the complete set of relevant parameter values is given in the Tab.~\ref{bplist}. 
\begin{table}
\begin{center}
\begin{tabular}{|l|c|c|}
\hline
Parameter&BP1&BP2\\
\hline
$m_{h^0}$&400 GeV&400 GeV\\
$m_{H^0}$&180 GeV&350 GeV\\
$m_{H^+}$& 220 GeV&350 GeV\\
$m_{A^0}$&30 GeV&40 GeV\\
$\tan\beta$&0.2&30\\
\hline
\hline
Branching ratio (\%)&BP1&BP2\\
\hline
$A^0\rightarrow b\overline{b}$&86&90\\
$A^0\rightarrow \tau^+\tau^-$&10&10\\
$H^0\rightarrow ZA^0$&$\sim$100&63\\
$H^\pm\rightarrow W^\pm A^0$&$\sim$100&79\\
$h^0\rightarrow H^+ H^-$&20&--\\
$h^0\rightarrow H^0 H^0$&10&--\\
\hline
\end{tabular}
\caption{Parameters values for the M2HDM that define the BP1 and BP2 benchmark points. The branching ratios relevant for the analyses presented in the following are also given. For simplicity, the small triplet mass splitting has been neglected in the BP2 case.\label{bplist}}
\end{center}
\end{table}

\subsection{Event simulation, reconstruction and selection cut definitions}

Signal and background events have been simulated using the generic 2HDM model of the tree-level matrix-element based event generator {\tt MadGraph/MadEvent} v4.4 \cite{Alwall:2007st,Alwall:2008pm}. The parameters of the model have been calculated using the {\tt TwoHiggsCalc} calculator \cite{Alwall:2007st}. The PDF set used is CTEQ6L \cite{Pumplin:2002vw} and the factorisation ($\mu_F$) and renormalisation ($\mu_R$) scales are evaluated on an event-by-event basis using relation
\begin{equation}
\mu^2_F=\mu^2_R=(M^2_{max}+\sum_j P_T^2)
\end{equation}
where $M_{max}$ is the larger mass among the final state particles and $j$ runs over the visible particles.

The showering/hadronisation phase, as well as the decay of unstable SM particles, are simulated using {\tt Pythia} 6.4 \cite{Sjostrand:2006za}.

In order to take into account the efficiency of event selection under realistic experimental conditions, the fast detector simulator {\tt Delphes} \cite{XavSev} has been used. Characteristics of the simulated detector, \ie , its geometry, granularity and resolution, are close to those associated with the ATLAS and CMS detectors. The tracker is assumed to reconstruct tracks within $|\eta|<2.5$ with a 100\% efficiency and the calorimeters cover a pseudo-rapidity region up to $|\eta|<3$ with an electromagnetic and hadronic tower segmentation of $\Delta\eta\sim 0.1$ and $\Delta\phi\sim 0.1$. The energy of each quasi stable particle is summed up in the corresponding calorimeter tower. The resulting energy is then smeared according to resolution functions assigned to the electromagnetic calorimeter (EC) and the hadronic calorimeter (HC) parameterized by:
\begin{eqnarray}
\frac{\sigma_{EC}}{E}&=&0.005+\frac{0.25}{E}+\frac{0.05}{\sqrt{E}}\\
\frac{\sigma_{HC}}{E}&=&0.05+\frac{1.5}{\sqrt{E}}.
\end{eqnarray}
where the energy $E$ is expressed in GeV.

The acceptance criteria are summarized in the Tab.~\ref{acctab}. For the lepton, we demand a tight isolation criterion: the number of additional tracks with $P_T>$ 1 GeV (denoted $N_{tracks}^{cone}$) present in a cone $\Delta R\equiv\sqrt{\Delta\eta^2+\Delta\phi^2}=0.3$ centered on the lepton track must be either 0 or 1. 
In so doing, we accommodate cases where collinear hard leptons are produced.
The jets are reconstructed using only the calorimeter towers, through the SISCone algorithm, as defined in the FastJet package \cite{Cacciari:2005hq} and implemented in {\tt Delphes}. Unless stated explicitly, a cone size radius of $0.7$ is applied for the jet algorithm. 
The $b$-tagging efficiency is assumed to be 40$\%$ for all $b$-jets, independently of their transverse momentum, with a fake rate of 1\% (10\%) for light (charm) jets. Finally, the total missing transverse energy $\slashed{E}_T$ is reconstructed using information from the calorimetric towers and muon candidates only. 
\begin{table}
\begin{center}
\begin{tabular}{|l|c|c|}
\hline
Final state&$|\eta^{max}|$&$P_T^{min}$ (GeV)\\
\hline
$e,\mu$&2.4&5\\
jets&3&40\\
$b$-jets&2.5&40\\
\hline
\end{tabular}
\caption{Acceptance of the different final states in the simulated detector.\label{acctab}}
\end{center}
\end{table}

In order to avoid repetitions in the forthcoming analyses, we define here a set of cuts:
\begin{itemize}
\item $A(l_i^n,j^m)$:  $m$ jets and $n$ leptons (electrons or muons) are required in the acceptance region with the isolation corresponding to $N_{tracks}^{cone}\leqslant i$.
\item $C_{nZ}$: $n$ $Z$ boson(s) are reconstructed from lepton kinematics. Lepton candidates fulfilling the acceptance cuts must have the same flavour, opposite charges, and a $P_T> 10$ GeV to reduce the amount of leptons from $B$ meson semi-leptonic decays. A $Z$ boson is then reconstructed if the di-lepton invariant mass lays in a 10 GeV mass window around the $Z$ mass.
\item $C_{b}$: at least one of the jet passing the acceptance cuts is $b$-tagged.
\item $C_A(l_1,l_2)$: The two leptons $l_1$ and $l_2$ have different flavours and opposite charges, belong to the same $\Delta R<1.2$ cone, and have an invariant mass smaller than 25 GeV.
\end{itemize}

In the following, we present simple strategies that can lead to promising Signal-over-Background ($S/B$) ratios. 
Our purpose is to illustrate the new possibilities that open up in the M2HDM and motivate more detailed studies. 
To this aim detailed information on the efficiencies and the visible cross sections are given.  The possibility for additional, more sophisticated, selection methods is also briefly addressed. 


\subsection{$b\overline{b}\rightarrow H^0\rightarrow Z A^0$}\label{bbza}

In a type II 2HDM (e.g., the MSSM scalar sector), the cross section of $b\overline{b}\rightarrow H^0$ is enhanced as $\tan\beta$ increases. This process has been shown to offer a promising discovery channel at the LHC when the Higgs boson decays into a $\tau^+\tau^-$ pair (e.g., see Ref.~\cite{Djouadi:2005gj} and references therein). For the mass spectrum defined by the benchmark point BP2 of the M2HDM, a particularly interesting decay mode is $H^0\to Z A^0$ (see Fig.~\ref{fig:feynsig1}).
 \begin{figure}[htbp]
\begin{center}
\begin{feynartspicture}(200,100)(1,1)
\FADiagram{}
\FAProp(-0.,19.5)(6.5,12.)(0.,){/Straight}{1}
\FALabel(2.5199,17.52)[bl]{$b$}
\FAProp(6.5,12.)(13.5,12.)(0.,){/ScalarDash}{0}
\FALabel(10.,13.02)[b]{$H^0$}
\FAProp(13.5,12.)(20.,19.5)(0.,){/ScalarDash}{0}
\FALabel(15.98,16.52)[br]{$A^0$}
\FAProp(20.,3.5)(13.5,12.)(0.,){/Sine}{0}
\FALabel(17.52,8.52)[bl]{$Z$}
\FAProp(-0.,3.5)(6.5,12.)(0.,){/Straight}{1}
\FALabel(2.6972,8.0596)[br]{$\overline{b}$}
\end{feynartspicture}
\caption{Feynman diagram for the $b\overline{b}\to H^0 \to ZA^0$ process.}
\label{fig:feynsig1}
\end{center}
\end{figure}

The production cross section of $b\overline{b}\rightarrow H^0$ is about 15 pb when normalized to the NNLO value reported in \cite{Harlander:2003ai}. Thanks to this large value, a wide range of interesting  final state signatures  can be considered. The cases where the $A^0$ boson provides two $b$-jets or $\tau$-jets are characterised by large rates, yet call for advanced identification and reconstruction techniques that need a dedicated experimental effort. For the sake of illustration we consider here the decay $A^0\rightarrow \tau^+\tau^-\rightarrow e^\pm\mu^\mp$ and $Z$ leptonic decays. Requiring different flavors in the $A^0$ decay considerably suppresses backgrounds with  $Z/\gamma^*\rightarrow e^+e^-$ or $\mu^+\mu^-$. 

The selected final state contains the three same flavour leptons $e^+e^- e^\pm \mu^\mp+ \slashed{E}_T$ and  $\mu^+\mu^- \mu^\pm e^\mp+ \slashed{E}_T$ .  
Such a multi-lepton final state is extremely clean and does not suffer from jet reconstruction uncertainties. The relevant backgrounds are $Z(Z/\gamma^*\rightarrow \tau\tau)$, $t\overline{t}Z$ and $W^+W^-Z$ with the decay of $W$ and $Z$ bosons into $e,\mu,\tau$ and $\tau\rightarrow e,\mu +\slashed{E}_T$. The cross sections for the signal and background processes after all decays are given in Tab.~\ref{xsecbbza}.
\begin{table}[ht]
\begin{center}
\begin{tabular}{|l|l|c|}
\hline 
Process& Decay (MC)& $\sigma\times BR$(fb)\\
\hline
\hline
$ZA$  & $ (Z\rightarrow l^+l^-)(A\rightarrow \tau\tau\rightarrow e^\pm\mu^\mp\slashed{E}_T)$  & 4.2 \\
 $Z(Z/\gamma)$          & $(Z\rightarrow \tilde{l}^+\tilde{l}^-)(Z/\gamma^*\rightarrow \tau\tau\rightarrow l^+l'^- \slashed{E}_T)$                    & 10\\
 $t\overline{t}Z$ & $ (t\rightarrow \tilde{l}^+ b\slashed{E}_T))(\overline{t}\rightarrow \tilde{l}^-\overline{b}\slashed{E}_T)) (Z\rightarrow \tilde{l}^+\tilde{l}^-)$     & 3.5 \\
 $W^+W^-Z$  &  $ (W^+\rightarrow \tilde{l}^+\slashed{E}_T)(W^-\rightarrow \tilde{l}^-\slashed{E}_T)(Z\rightarrow \tilde{l}^+\tilde{l}^-)$            & 0.4\\
 \hline
\end{tabular}
\caption{Cross sections of signal $ZA^0$ and backgrounds processes taking into account the leptonic final state considered in the analysis. The notation $l$ includes only $e$ and $\mu$, whereas $\tilde{l}$ also contains $\tau$ decaying into $e$ or $\mu$. All cross sections correspond to the final states in the second column.
\label{xsecbbza}}
\end{center}
\end{table}
The selection proceeds as follows. The acceptance cut $A(l_1^4)$ is applied, followed by the $C_{1Z}$ cut. The same flavour opposite sign leptons paired whose mass is the closest to the $Z$ mass is retained. The two remaining leptons $l_1$ and $l_2$ are then assumed to come from the light and boosted pseudoscalar, and forced to satisfy $C_A(l_1,l_2)$. The relative and total efficiencies, as well as the visible cross sections for all processes listed in Tab.~\ref{xsecbbza} are reported in Tab.~\ref{bbzaeff}. 
\begin{table}[h]
\begin{center}
\begin{tabular}{|lc|c|c|c|c|c|c|}
\hline
&&$ZA$&$Z(Z/\gamma^*)$&$t\bar{t}Z$\\
\hline
\hline
$A(l^4_1)$&($\%$)&51&18&42\\
$C_Z$&($\%$)&74&63&60\\
$C_A$&($\%$)&85&3.6&3.3\\
\hline
$\epsilon_{tot}$&$(\%)$&32&0.39&0.84\\
$\sigma_{vis}$&$(fb)$&1.5&0.039&0.029\\
\hline
\end{tabular}
\caption{Relative efficiencies of the considered cuts, together with the total efficiencies after all cuts and corresponding visible cross sections for signal and background processes. The WWZ process is omitted since its visible cross section is four orders of magnitude smaller than that of the signal.
\label{bbzaeff}}
\end{center}
\end{table}

As can be clearly seen, this channel is very promising: the $S/B$ is high enough so that an excess over the SM could be identified after a few inverse femtobarns of integrated luminosity. Note that the ($Z\rightarrow l^+l^-$) + jets background has also been considered due to its very large cross section (${\cal O}$(nb)), the possibility for jets to produce fake electrons, and the possible presence of leptons from heavy meson decays. An inclusive sample of $10^6$  events  was generated using the matching procedure \cite{Alwall:2008qv} and no event has passed the isolation cuts. This background is therefore neglected. However a more detailed study should be performed with a more realistic detector simulation and event reconstruction.

\begin{figure}[h!]
      \centering \includegraphics[width=0.98\textwidth]{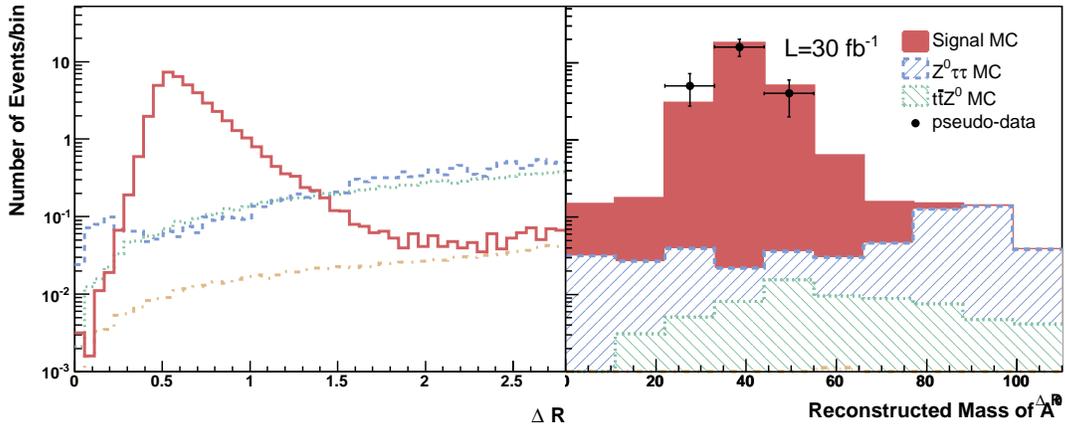}
      \caption{Left: the number of events in function of the distance $\Delta R$ between the two leptons not assigned to the $Z$, after applying the $A(l_1^3)$ and $C_Z$ cuts. Right: the number of events in function of the reconstructed mass of $A^0$ using the leptons 4-vectors and the missing $E_T$ after applying the additional $C_A$ cut (which require $\Delta R <  1.2$). Both figures correspond to an integrated luminosity of 30 fb$^{-1}$. The markers show a hypothetical event excess for this luminosity.\label{Amass}}
\end{figure}

Besides a pure counting experiment a more exclusive study can also be attempted. The mass of the two neutral resonances could be measured with an accuracy depending mostly on the $\slashed{E}_T$ reconstruction quality. In the signal, the main source of missing transverse energy originates from the $\tau$'s. If the direction of the $\slashed{E}_T$ is required to lay between the transverse position of the two leptons $l_1$ and $l_2$ and the condition  $\slashed{E}_T>50$ GeV imposed, then a proper reconstruction of the invariant mass $m_{A^0}$ can be achieved (see Fig.~\ref{Amass}). A substantial improvement of the $S/B$ ratio is also gained. Finally, the $H^0$ mass can be estimated from the $A^0$ and $Z$ boson 4-vectors.


\subsection{$g(b/\overline{b})\rightarrow (t/\overline{t})H^\pm \rightarrow W^- (b/\overline{b}) W^+ A^0$}
\label{chargedhiggs}
The previous analysis shows that, for the benchmark point BP2, a discovery could be made after a few inverse femtobarns of integrated luminosity together with the identification of two neutral Higgs bosons $H^0$ and $A^0$. However, in order to fully determine the structure of an extended scalar sector, it is also crucial to observe a charged Higgs boson. In the M2HDM, we expect it to be nearly degenerate in mass with $H^0$ as a consequence of the twisted custodial symmetry.

The associated production of a charged Higgs with a top quark, $g(b/\overline{b})\rightarrow (t/\overline{t})H^\pm$  (see Fig.~\ref{fig:feynsig2}), is in general considered as a challenging channel at the LHC. The discovery potential  strongly depends  on $\tan\beta$, the mass of the charged Higgs boson and the considered decay mode. However, at variance with models such as the MSSM, the M2HDM offers the possibility for the $H^\pm$ to decay into $W^\pm A^0$. Its observation would therefore be a very strong evidence that the scalar sector originates from the M2HDM. We consider this possibility in the benchmark point BP2. To normalize the expected signal, we use the NLO prediction for the charged Higgs production cross section from Ref.~\cite{Plehn:2002vy}, \ie , 465 fb. 
\begin{figure}[htbp]
\begin{center}
\begin{feynartspicture}(250,125)(1,1)

\FADiagram{}

\FAProp(10.,5.)(-0.,5.)(0.,){/Straight}{-1}
\FALabel(5.3771,5.7922)[b]{$\overline{b}$}
\FAProp(10.,13.5)(10.,5.)(0.,){/Straight}{-1}
\FALabel(9.28,9.25)[r]{$\overline{t}$}
\FAProp(14.,15.)(10.,13.5)(0.,){/Straight}{-1}
\FALabel(11.6487,14.7063)[b]{$\overline{t}$}
\FAProp(-0.,13.5)(10.,13.5)(0.,){/Cycles}{0}
\FALabel(5.,12.68)[t]{}
\FAProp(20.,18.)(14.,15.)(0.,){/Straight}{-1}
\FALabel(18.48,18.27)[br]{$\overline{b}$}
\FAProp(14.,15.)(20.,12.5)(0.,){/Sine}{0}
\FALabel(18.5,14.02)[b]{$W^-$}
\FAProp(10.,5.)(14.,5.)(0.,){/ScalarDash}{0}
\FALabel(11.9999,4.1798)[t]{$H^+$}
\FAProp(14.,5.)(20.,8.)(0.,){/ScalarDash}{0}
\FALabel(18.48,7.52)[br]{$A^0$}
\FAProp(14.,5.)(19.5,0.5)(0.,){/Sine}{0}
\FALabel(18.52,3.02)[bl]{$W^+$}
\end{feynartspicture}
\caption{Representative Feynman diagram for the $g\overline{b}\to \overline{t}H^+ \to W^+W^- \overline{b} A^0$ process.}
\label{fig:feynsig2}
\end{center}
\end{figure}

As in the previous analysis, we focus exclusively on the decay $A^0\rightarrow\tau^+\tau^-$ where the $\tau^+\tau^-$ pair decays into $e^\pm\mu^\mp$. Despite the fact the total signal cross section is reduced by almost two orders of magnitude compared to the $A^0\to b\overline{b}$ case, a strong reduction of the background is foreseen if one of the $W$ bosons decays leptonically. The considered final state is therefore $l^\pm jjb e^\pm \mu^\mp+\slashed{E}_T$. If the light quark pair comes from the $W$ boson produced in the charged Higgs decay, the resulting jets tend to be collinear due to the large boost. As a consequence, they might not be resolved but merged into a single ``large'' jet (noted $J$) by the reconstruction algorithms. We include both possibilities.

The relevant backgrounds are $W^+W^-W^\pm jj$, $t\overline{t}(Z/\gamma^*)$, $W(Z/\gamma^*)jj$, $Z(Z/\gamma^*)jj$, and $tW (Z/\gamma^*)$, with $j$ standing for all light and $b$ quarks. The cross sections for the signal and the considered background processes, as well as the corresponding final states, are summarized in Tab.~\ref{tabthpm}. For $W^\pm/Z+\tau\tau+jj$ and $W^\pm W^+W^-jj$, the jets are initially produced with a minimal $P_T$ of 10 GeV, a maximal pseudo-rapidity of 5, and an angular separation of $\Delta R(jj) > 0.1$ for the firsts and $\Delta R(jj) > 0.2$ for the latter. The details of the decay modes and the corresponding rates are shown in Tab.~\ref{tabthpm}.
\begin{table}[ht]
\begin{center}
\begin{tabular}{|l|l|c|}
\hline 
Process & Decay (MC) & $\sigma\times BR$ (fb)\\
\hline
\hline
\multirow{2}{2cm}{$(\overline{t}/t)H^\pm$}  & $ ((t/\overline{t})\rightarrow \tilde{l}^\mp b \slashed{E}_T)(H^\pm\rightarrow (W^\pm\rightarrow jj) (A\rightarrow\tau\tau\rightarrow e^\pm \mu^\mp \slashed{E}_T))(b)$& 0.75 \\
 & $ ((t/\overline{t})\rightarrow jj b)(H^\pm\rightarrow (W^\pm\rightarrow \tilde{l}^\pm \slashed{E}_T) (A\rightarrow\tau\tau\rightarrow e^\pm \mu^\mp \slashed{E}_T))(b)$ & \\
 $t\overline{t}(Z/\gamma^*)$ & $ (t\rightarrow incl.)(\overline{t}\rightarrow incl.) (Z/\gamma^*\rightarrow \tilde{l}'^+\tilde{l}''^-\slashed{E}_T)$     & 4.5 \\
 $W(Z/\gamma^*)jj$     & $ (W^\pm\rightarrow \tilde{l}^\pm\slashed{E}_T)(Z/\gamma^*\rightarrow \tilde{l}'^+\tilde{l}''^-\slashed{E}_T)jj$                & 48\\
 $Z(Z/\gamma)jj$          & $(Z\rightarrow \overline{l}^+\overline{l}^-)(Z/\gamma^*\rightarrow \tilde{l}'^+\tilde{l}''^-\slashed{E}_T)jj$                    & 10\\
 $(t/\overline{t})W(Z/\gamma^*)$  &  $ (t\rightarrow incl.)(l^\pm \rightarrow \tilde{l}'^+\tilde{l}''^-\slashed{E}_T)$            & 0.6\\
  $W^+W^-W^\pm jj$           & $(W^\pm\rightarrow \tilde{l}^\pm\slashed{E}_T)(W^+\rightarrow \tilde{l}^+\slashed{E}_T)(W^-\rightarrow \tilde{l}^-\slashed{E}_T)jj$                                           & 13\\
 \hline
\end{tabular}
\caption{Cross sections of the signal $gb\rightarrow tH^\pm \rightarrow W^- b W^+ A^0$ and the relevant background processes, taking into account leptonic and jet final states considered in the analysis. The notation $\tilde{l}$ means that the three flavour of leptons are taken into account, and the tau leptons decay into $e$ or $\mu$. On the contrary, $\overline{l}$ means that the tau leptons decay inclusively ($Z/(Z/\gamma^*)$ case). All the quoted cross sections correspond to the final states in the second column.
\label{tabthpm}}
\end{center}
\end{table}

\begin{figure}[h!]
\begin{center}
\includegraphics[width=0.98\textwidth]{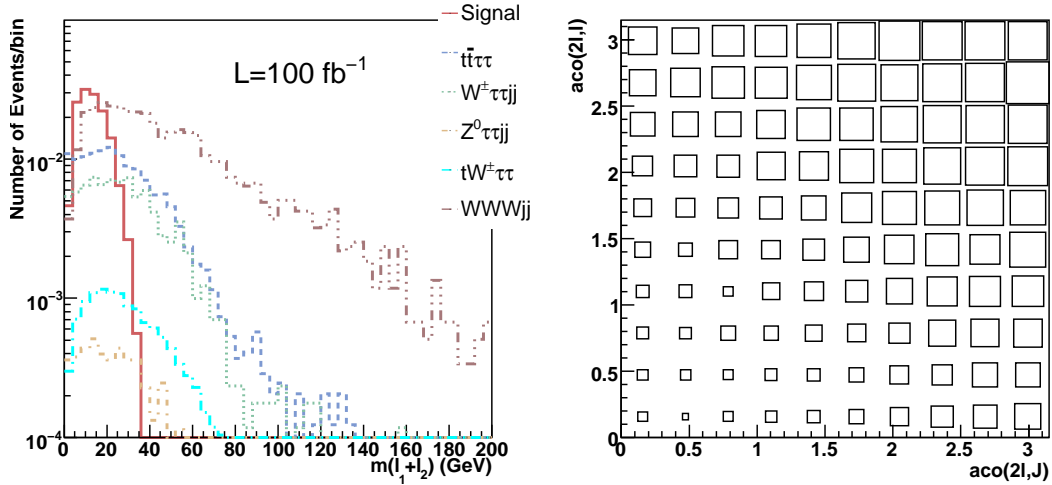}
\caption{Left: invariant mass distribution of the two closest leptons (with different charges and flavours) for both signal and background events. Right: two dimensional distribution of events after $A(l_1^3,j^2)$, $C_b$, $\overline{C}_Z$ and $C_A$ cuts, as a function of the acoplanarity between the sum of collinear leptons and the hardest non $b$-tagged jet acoplanarity ${\rm aco}(2l,J)$, and the third lepton ${\rm aco}(2l,l)$  .
\label{plotvarthpm}}
\end{center}
\end{figure}

In order to increase the $S/B$ ratio, the acceptance cut $A(l_1^3,j^2)$ and the $C_b$ cut are applied. These are followed by a veto on the presence of a $Z$ boson  $\overline{C}_Z$. The two closest leptons ($l_{1}$, $l_{2}$) with opposite charges and different flavours are assume to come from the light and boosted pseudoscalar Higgs boson $A^0$, and therefore are required to satisfy the $C_A(l_1,l_2)$ cut. The relevance of this last cut is illustrated in Fig.~\ref{plotvarthpm}, where the left-hand side plot shows the di-lepton invariant mass after applying all cuts except $C_A(l_1,l_2)$. 
The relative and total efficiencies, as well as the visible cross sections for all processes listed in Tab.~\ref{tabthpm}, are reported in Tab.~\ref{tabeffthpm}.
\begin{table}[h]
\begin{center}
\begin{tabular}{|lc|c|c|c|c|c|c|c|}
\hline
&&$tH^\pm$&$t\overline{t}(Z/\gamma^*)$&$W(Z/\gamma^*)jj$&$Z(Z/\gamma^*)jj$&$tW (Z/\gamma^*)$&$W^\pm W^+W^-jj$\\
\hline
\hline
$A(l_1^3,j^2)$&$(\%)$&35&16&5.5&3.6&14&21\\
$C_b(\%)$&$(\%)$&39&48&6&6.3&39&49\\
$\overline{C}_Z(\%)$&$(\%)$&98&98&91&60&98&95\\
$C_A(\%)$&$(\%)$&84&15&19&12&11&6.5\\
\hline
$\epsilon_{tot}$&$(\%)$&11&1.1&0.056&0.017&0.61&0.64\\
$\sigma_{vis}$&$(fb)$&0.083&0.051&0.027&0.0017&0.0037&0.083\\
\hline
\end{tabular}
\caption{Relative efficiencies of the various cuts together with the total efficiencies after all cuts and corresponding visible cross sections for signal and background processes.
\label{tabeffthpm}}
\end{center}
\end{table}

A rather low visible signal cross section confirms that this channel is also very challenging with the unusual $H^\pm\rightarrow W^\pm A^0$ decays. However, the $S/B$ ratio of order ${\cal O}(1)$ leaves some hope that a charged Higgs could still be discovered after a large integrated luminosity ($\sim 300$ fb$^{-1}$). In any case it should be kept in mind that the benchmark point BP2 is not the most optimistic scenario: a lighter $H^\pm$ associated with a larger $\tan\beta$ would sizably increase the production cross section. 

In addition, more exclusive discriminant variables could be used to exploit further the characteristics of the typical topology. As an example, let us consider the fact that the heaviest particle in the process is the charged Higgs boson with at least twice the mass of the top quark. As a result, it is typically produced with a small transverse momentum, giving acoplanar $W^\pm$ and $A^0$ bosons with large boost. 
This acoplanarity ($\Delta\phi$ between considered final states) can be estimated from the two collinear lepton $e^\pm\mu^\mp$  together with the decay products of the $W$ originating from the charged Higgs. This decay product is either the third lepton, or the ``large" jet $J$ if the $W$ boson from the charged Higgs decays hadronically. Since the two topologies are a priori not known, the two acoplanarity definitions (resp. ${\rm aco}(2l,l)$ and ${\rm aco}(2l,J)$) are built for each event. The distribution of signal events with respect to these two variables is illustrated on the right plot of Fig.~\ref{plotvarthpm}. As the distribution of background events is much more uniform in this plane, an enhancement of the $S/B$ ratio of around 10$\%$ can be achieved if a cut ${\rm aco}(2l,l)+{\rm aco}(2l,J) > 3$  is applied.


\subsection{$gg\to h^0\to H^0 H^0\to ZA^0 ZA^0$}

\begin{figure}[htbp]
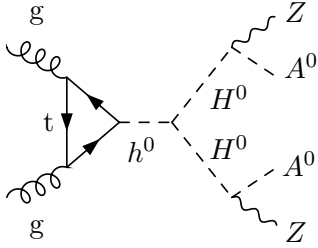

\begin{center}
\begin{feynartspicture}(250,125)(1,1)
\FADiagram{}
\FAProp(4.,13.)(4.,7.)(0.,){/Straight}{1}
\FALabel(3.18,10.)[r]{t}
\FAProp(4.,7.)(0.,5.)(0.,){/Cycles}{0}
\FALabel(1.52,3.48)[tl]{g}
\FAProp(0.,15.)(4.,13.)(0.,){/Cycles}{0}
\FALabel(1.52,16.52)[bl]{g}
\FAProp(7.5,10.)(4.,13.)(0.,){/Straight}{1}
\FALabel(5.52,12.52)[bl]{}
\FAProp(4.,7.)(7.5,10.)(0.,){/Straight}{1}
\FALabel(5.52,7.48)[tl]{}
\FAProp(7.5,10.)(11.,10.)(0.,){/ScalarDash}{0}
\FALabel(9.,8.98)[t]{$h^0$}
\FAProp(11.,10.)(15.,15.)(0.,){/ScalarDash}{0}
\FALabel(13.52,12.48)[tl]{$H^0$}
\FAProp(11.,10.)(15.,5.)(0.,){/ScalarDash}{0}
\FALabel(13.52,7.52)[bl]{$H^0$}
\FAProp(15.,15.)(18.,17.)(0.,){/Sine}{0}
\FALabel(18.52,17.98)[tl]{$Z$}
\FAProp(15.,15.)(18.,13.)(0.,){/ScalarDash}{0}
\FALabel(18.4696,12.9618)[bl]{$A^0$}
\FAProp(15.,5.)(18.,3.)(0.,){/Sine}{0}
\FALabel(18.52,2.02)[bl]{$Z$}
\FAProp(15.,5.)(18.,7.)(0.,){/ScalarDash}{0}
\FALabel(18.3456,8.0304)[tl]{$A^0$}
\end{feynartspicture}
\caption{Feynman diagram for  the $gg\to h^0\to H^0 H^0\to Z A^0 Z A^0$ process.}
\label{fig:feynsig3}
\end{center}
\end{figure}

In the context of SM Higgs searches  at the LHC, it has been shown that the discovery of $h^0$ at high mass could be achieved after a few inverse femtobarns, notably with the observation of $h^0\rightarrow ZZ$ or $W^+ W^-$ decays \cite{:1999fq,Ball:2007zza}. A deviation from the expected visible cross section could reveal  the presence of additional decays of the Higgs boson such as those predicted by the M2HDM. In this context, it is interesting to study the process $gg\to h^0\to H^0 H^0$  (see Fig.~\ref{fig:plotbr_1_h0}) since it benefits from a relatively large gluon fusion production cross section (around 10 pb at $m_{h^0}$ of 400 GeV, when using the NNLO prediction from \cite{Catani:2003zt}) and a sizable $h^0\to H^0 H^0$ branching ratio. This possibility is investigated assuming the benchmark point BP1 of the M2HDM and considering the process $gg\to h^0\to H^0 H^0\to Z A^0 Z A^0$. It also provides a clear  in the final state signature thanks to the presence of two $Z$ bosons (see Fig.~\ref{fig:feynsig3}),  required to decay into $e$ and $\mu$. 

Under this decay hypothesis, the total rate decreases considerably, such that only the main $A^0\to b\overline{b}$ decay mode can reasonably be retained. At the parton level the signal final state is then $l^+l^-l'^+l'^- b\overline{b}b\overline{b}$, with $l^\pm=e^\pm,\mu^\pm$. 
The $b$-quarks produced by the pseudoscalar $A^0$ are well separated ($\Delta R \gtrsim$ 1) but have a low average transverse momentum. In order to keep a good efficiency by merging the jets two-by-two for the signal, a large cone size radius ($\approx$ 1) is used. This leads to a final state with only main two jets in the final state. It should be noted, however, that with a heavier $A^0$ the individual detection of all four jets could be attempted.

The main backgrounds to be taken into account are $t\overline{t}Z$ and $ZZjj$. The process $gg\rightarrow h^0\rightarrow ZZ$ can be neglected as well as $W^+W^-Z$ because of their relatively low cross section and low probability to provide a $b$-tagged jet. For $ZZjj$ the jets are produced with a minimal $P_T$ of 20 GeV, a maximal pseudo-rapidity of 5 and a $\Delta R(jj) > 0.3$. The cross sections for the signal and background processes are given in Tab.~\ref{tblZAZAxsex}
\begin{table}[htbp]
\begin{center}
\begin{tabular}{|l|l|c|}
\hline 
Process & Decay (MC) & $\sigma$ (fb)\\
\hline
\hline
$ZAZA$  & $ (Z\rightarrow l^+ l^-)(Z\rightarrow l'^+l'^-) b\overline{b} b\overline{b}$  & 3.2 \\
 $ZZjj$          & $(Z\rightarrow l^+ l^-)(Z\rightarrow l'^+l'^-)jj$                    & 16\\
  $t\overline{t}Z$ & $ (t\rightarrow \tilde{l}^+ b\slashed{E}_T))(\overline{t}\rightarrow \tilde{l}^-\overline{b}\slashed{E}_T)) (Z\rightarrow \tilde{l}^+\tilde{l}^-)$     & 3.5 \\
 \hline
\end{tabular}
\caption{List of processes considered in the analysis of the $ZA^0ZA^0$ channel. The notation $l$ means that only electron and muons are considered. If the notation $\tilde{l}$ is used instead, all flavours are included  and the taus are decayed in $e$ or $\mu$.\label{tblZAZAxsex}}
\end{center}
\end{table}

In order to increase the ratio of the $S/B$ ratio, the acceptance cut $A(l_0^4,j^2)$ and the $C_b$ cut are applied. The efficiency of the $C_b$ is assumed to be the same as for single $b$ quark induced jets. We then apply $C_{2Z}$ cut, where the invariant mass of the two pairs of same-flavour leptons are the closest to the actual $Z$ mass. 
The relative and total efficiencies  for all processes listed in Tab.~\ref{tblZAZAxsex}, as well as their visible cross sections, are reported in Tab.~\ref{tabZAZA}. The visible cross section around 0.3 fb and a $S/B$ ratio close to 3 suggest that, using only the simple discriminant variables described here above, the evidence of such a signal could be reached with a total integrated luminosity smaller than 30 fb$^{-1}$.
\begin{table}[h]
\begin{center}
\begin{tabular}{|lc|c|c|c|c|c|c|c|}
\hline
&&$ZAZA$&$ZZjj$&$t\bar{t}Z$\\
\hline
\hline
$A(l^4_0,j^2)$&$(\%)$&27&11&18\\
$C_b$              &$(\%)$&50&7.9&54\\
$C_{2Z}$         &$(\%)$&72&75&4.1\\
\hline
$\epsilon_{tot}$&$(\%)$&9.6&0.63&0.4\\
$\sigma_{vis}$&$(fb)$&0.32&0.1&0.014\\
\hline
\end{tabular}
\caption{Relative efficiencies (in percent) for each cut presented in the text. Combined efficiencies and resulting visible cross sections after all cuts are also shown. \label{tabZAZA}}
\end{center}
\end{table}

\begin{figure}[t]
\begin{center}
\includegraphics[width=0.9\textwidth]{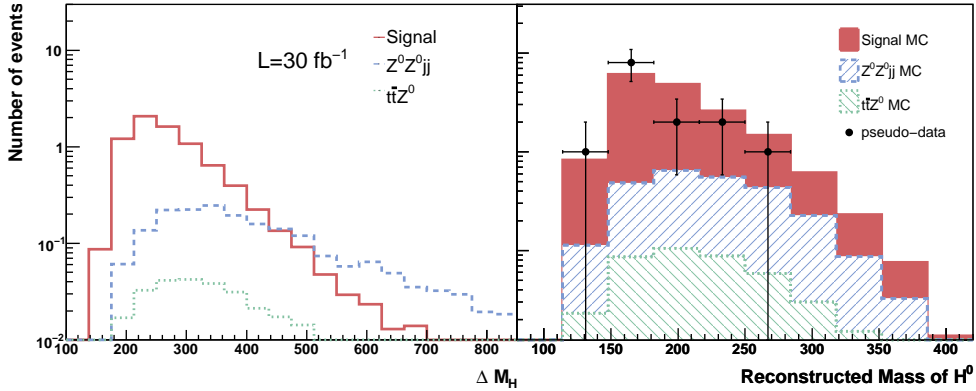}
\caption{Left: the difference $\Delta M_{H}$ between $m_{h^0}$ and the mean of $m_{H^0}$ after $A(l_0^4,j^2)$, $C_b$ and $C_{2Z}$ cuts. Right: invariant masses of the pseudoscalar $H^0$ bosons (two entries per event). The markers show a hypothetical event excess for this luminosity.
\label{mhmH}}
\end{center}
\end{figure}

After application of the reviewed ``standard'' cuts, the $S/B$ ratio could be further improved by applying a more sophisticated cut, taking advantage of the fact that invariant mass of the $h^0$ can, in principle, be fully reconstructed. First, the two $H^0$ masses are reconstructed, each from one $Z$ and one jet (which for the signal, has to be understood as the single jet induced by the $A^0$ boson decay), such that the difference between the two possible $H^0$ mass combinations must be minimal and smaller than 100 GeV. Furthermore, the mass difference between the $h^0$ candidate directly reconstructed from the sum of all 4-vectors of the four leptons and the two jets, and the mean of $H^0$ candidate masses, must be smaller than 400 GeV. The relevance of this cut is illustrated in Fig.~\ref{mhmH} where the left-hand side plot shows the distribution of signal and background events as a function of the difference between $m_{h^0}$ and the mean of $m_{H^0}$, after applying all other cuts. The $S/B$ ratio could be easily increased up to 5, to the price of a lower signal visible cross section (by roughly 30\%).

The distribution of the $H^0$ reconstructed mass for the events passing all cuts is also shown in Fig.~\ref{mhmH}. It illustrates the possibility to measure this parameter with the simple algorithm described above. The resolution could certainly be improved as well as the signal significance if a proper jet reconstruction with an optimal cone size and the tracker information were taken into account. The determination of the invariant mass of the pseudoscalar $A^0$ could also be attempted.

Note however that the conclusions of this section has to be interpreted with some caution since $b$-tagging efficiency and jet kinematics have to be re-evaluated when two soft and/or collinear $b$-induced jets are merged into one jet. This question can only be precisely addressed with a proper full simulation of detector effects.

\section{Conclusion}
We have examined a minimal extension of the SM scalar sector based on natural symmetries. Once a suitable mechanism is provided to generate the required triplet mass splitting, the proposed model (which we have dubbed Minimal 2HDM) may fulfill existing theoretical and experimental constraints, while giving rise to unexpected phenomenology at high energy colliders. 

Our starting point, the generic 2HDM, is a simple yet rich framework which may arise in the context of various BSM theories. Genuine SM properties give grounds for specific global symmetries, like the custodial and $CP$ ones, which may or may not be satisfied by new interactions. Their implementation in the context of the 2HDM sheds new light on their possible interplay and allows us to naturally introduce a ``twisted'' scenario. In this new scenario, the successful $\rho\approx 1$ phenomenological relation is ensured by the degeneracy of a pair of charged scalars ($H^\pm$) with a scalar ($H^0$), and not with a pseudoscalar ($A^0$) at variance with the usual case in the 2HDM.

Surprisingly enough, this seemly mild difference opens a window to novel and, to a large extent, unexplored phenomenology of the Higgs sector. Due to its vanishing coupling to a pair of gauge boson, the pseudoscalar state $A^0$ might have escaped the LEP II searches. The mass of charged Higgs boson, on the other hand, is already strongly constrained by indirect measurements. In the twisted scenario we propose, the large mass splitting between $(H^\pm,H^0)$ and $A^0$ is naturally reconciled with tight electroweak precision constraints. A SM Higgs boson mass larger than $\sim 200$ GeV may also be accommodated through small breaking of the custodial symmetry, driven for example by loop corrections.

We have illustrated how an inverted mass spectrum $m_{A^0} \ll m_{H^0}\lesssim m_{H^\pm} \ll m_{h^0}$, as compared for example to the typical MSSM scenario, with a relatively light $A^0$ (e.g., $m_{A^0}\approx 30$ GeV) could meet all the theoretical and present indirect and direct constraints. It could even appear as favoured by some precision measurement like the $S$ parameter or the muon anomalous magnetic moment. This particular mass spectrum leads to interesting new signal opportunities, mainly related to the opening up of new decay channels such as $h^0\to H^0 H^0, H^+H^-$, $H^\pm\to W^\pm A^0$ or $H^0\to Z A^0$. To our knowledge, these signals have been only partially covered by analyses in the context of more restricted BSM models like the NMSSM or the $CP$ violating MSSM. 

A first analysis, including a fast simulation of detector effects, has been performed for three characteristic channels at the LHC.  In our scenarios the SM-like Higgs boson is at  400 GeV and could be easily found in the $ZZ$ and $WW$ decay channels after a few inverse femtobarns of integrated luminosity.  At that point, the possibility of an extended scalar sector and in particular of a M2HDM with an inverted hierarchy should be explored.
Our results suggest that the neutral Higgs bosons $H^0$ and $A^0$ could be discovered and their mass measured already in the low luminosity phase of the LHC running. As a particularly illuminating channel we have proposed to look for $H^0$ production in $b\overline{b}$ fusion and its subsequent decay $H^0\to Z A^0$ with $A^0\to \tau^+ \tau^-$.  In this case a discovery could be achieved after only a few femtobarns of integrated luminosity. As an alternative or in addition to the above process, we have also studied the  $gg\to h^0\to H^0 H^0\to ZA^0 ZA^0$ channel with $A^0\to b\overline{b}$, in a Type I scenario. We find that despite its rather low cross section the signal could  be disentangled from the (extremely low) SM backgrounds and could provide a rather unique chance to test the inverted mass spectrum hypothesis. 
Imagining that three new scalars were observed, the search for the last component of the scalar doublet, \ie, the charged Higgs, would become the first priority.  Such a discovery, however, will probably be as challenging as in other more conventional scenarios. As a promising and unconventional channel we have considered the associated production of a charged Higgs with a single top quark, with the decays $H^\pm \to W^\pm A^0$ and $A^0\to\tau^+\tau^-$.  Our analysis shows that the rather small signal rate over a large three-$W$'s background would probably need a very large integrated luminosity before being significant.

The perspectives for extending the present work can be categorised along two main directions. On the theoretical side, the twisted version of the custodial and $CP$ symmetries can probably be generalised to other sectors of the theory, like the Yukawa sector, or to more complex scalar sector extensions, \ie , models with more than two Higgs doublets or involving higher representations. Applications to more ``ambitious'' BSM scenarios like Technicolor-inspired models could also be envisaged. 


On the experimental side, more detailed studies would certainly be welcome. Our preliminary results reveal some unusual and specific challenges characteristic of proposed scenarios. Among others, the identification of highly collinear lepton or $b$-jet pairs is particularly interesting and appears as a recurrent requirement for several processes. For the latter case various innovative solutions to these problems may exist or have already been proven to exist (e.g. in \cite{Butterworth:2008sd}) and some of them might be applied to the study of the M2HDM in a near future.

\acknowledgments{MH thanks A. Djouadi, H. Haber and C. Wagner for interesting comments during various presentations of the results presented in this article. This work was supported by the Institut Interuniversitaire des Sciences Nucl\'eaires and by the Belgian Federal Office for Scientific, Technical and Cultural Affairs through the Interuniversity Attraction Pole P6/11. It is also part of the research program of the ``Stichting voor Fundamenteel Onderzoek der Materie (FOM)", which is financially supported by the ``Nederlandse organisatie voor Wetenschappelijke Onderzoek (NWO)".}

\bibliography{2HDM}

\providecommand{\href}[2]{#2}\begingroup\raggedright\begin{thebibliography}{10}

\bibitem{Higgs:1964pj}
P.~W. Higgs, {\it Broken symmetries and the masses of gauge bosons},  {\em
  Phys. Rev. Lett.} {\bf 13} (1964) 508--509.

\bibitem{Englert:1964et}
F.~Englert and R.~Brout, {\it Broken symmetry and the mass of gauge vector
  mesons},  {\em Phys. Rev. Lett.} {\bf 13} (1964) 321--322.

\bibitem{Guralnik:1964eu}
G.~S. Guralnik, C.~R. Hagen, and T.~W.~B. Kibble, {\it Global conservation laws
  and massless particles},  {\em Phys. Rev. Lett.} {\bf 13} (1964) 585--587.

\bibitem{Veltman:1997nm}
M.~J.~G. Veltman, {\it Reflections on the higgs system}, . CERN-97-05.

\bibitem{Sikivie:1980hm}
P.~Sikivie, L.~Susskind, M.~B. Voloshin, and V.~I. Zakharov, {\it Isospin
  breaking in technicolor models},  {\em Nucl. Phys.} {\bf B173} (1980) 189.

\bibitem{Glashow:1976nt}
S.~L. Glashow and S.~Weinberg, {\it Natural conservation laws for neutral
  currents},  {\em Phys. Rev.} {\bf D15} (1977) 1958.

\bibitem{Gerard:2007kn}
J.~M. Gerard and M.~Herquet, {\it {A twisted custodial symmetry in the
  two-Higgs-doublet model}},  {\em Phys. Rev. Lett.} {\bf 98} (2007) 251802,
  [\href{http://xxx.lanl.gov/abs/hep-ph/0703051}{{\tt hep-ph/0703051}}].

\bibitem{Dermisek:2008id}
R.~Dermisek, {\it {Light CP-odd Higgs and Small $\tan \beta$ Scenario in the
  MSSM and Beyond}},  \href{http://xxx.lanl.gov/abs/0806.0847}{{\tt
  0806.0847}}.

\bibitem{Dermisek:2008uu}
R.~Dermisek and J.~F. Gunion, {\it {Many Light Higgs Bosons in the NMSSM}},
  {\em Phys. Rev.} {\bf D79} (2009) 055014,
  [\href{http://xxx.lanl.gov/abs/0811.3537}{{\tt 0811.3537}}].

\bibitem{DiazCruz:1992uw}
J.~L. Diaz-Cruz and A.~Mendez, {\it {Vacuum alignment in multiscalar models}},
  {\em Nucl. Phys.} {\bf B380} (1992) 39--50.

\bibitem{Ferreira:2004yd}
P.~M. Ferreira, R.~Santos, and A.~Barroso, {\it {Stability of the tree-level
  vacuum in two Higgs doublet models against charge or CP spontaneous
  violation}},  \href{http://xxx.lanl.gov/abs/hep-ph/0406231}{{\tt
  hep-ph/0406231}}.

\bibitem{Branco:1999fs}
G.~C. Branco, L.~Lavoura, and J.~P. Silva, {\em {CP violation}}.
\newblock Oxford, UK: Clarendon, 1999.

\bibitem{Ginzburg:2004vp}
I.~F. Ginzburg and M.~Krawczyk, {\it {Symmetries of two Higgs doublet model and
  CP violation}},  {\em Phys. Rev.} {\bf D72} (2005) 115013,
  [\href{http://xxx.lanl.gov/abs/hep-ph/0408011}{{\tt hep-ph/0408011}}].

\bibitem{Davidson:2005cw}
S.~Davidson and H.~E. Haber, {\it {Basis-independent methods for the
  two-Higgs-doublet model}},  {\em Phys. Rev.} {\bf D72} (2005) 035004,
  [\href{http://xxx.lanl.gov/abs/hep-ph/0504050}{{\tt hep-ph/0504050}}].

\bibitem{Haber:2006ue}
H.~E. Haber and D.~O'Neil, {\it {Basis-independent methods for the
  two-Higgs-doublet model. II: The significance of tan(beta)}},  {\em Phys.
  Rev.} {\bf D74} (2006) 015018,
  [\href{http://xxx.lanl.gov/abs/hep-ph/0602242}{{\tt hep-ph/0602242}}].

\bibitem{LopezHonorez:2006gr}
L.~Lopez~Honorez, E.~Nezri, J.~F. Oliver, and M.~H.~G. Tytgat, {\it {The inert
  doublet model: An archetype for dark matter}},  {\em JCAP} {\bf 0702} (2007)
  028, [\href{http://xxx.lanl.gov/abs/hep-ph/0612275}{{\tt hep-ph/0612275}}].

\bibitem{Toussaint:1978zm}
D.~Toussaint, {\it Renormalization effects from superheavy higgs particles},
  {\em Phys. Rev.} {\bf D18} (1978) 1626.

\bibitem{Chankowski:2000an}
P.~Chankowski {\em et~al.}, {\it {Do precision electroweak constraints
  guarantee e+ e- collider discovery of at least one Higgs boson of a two Higgs
  doublet model?}},  {\em Phys. Lett.} {\bf B496} (2000) 195--205,
  [\href{http://xxx.lanl.gov/abs/hep-ph/0009271}{{\tt hep-ph/0009271}}].

\bibitem{Barbieri:2006dq}
R.~Barbieri, L.~J. Hall, and V.~S. Rychkov, {\it {Improved naturalness with a
  heavy Higgs: An alternative road to LHC physics}},  {\em Phys. Rev.} {\bf
  D74} (2006) 015007, [\href{http://xxx.lanl.gov/abs/hep-ph/0603188}{{\tt
  hep-ph/0603188}}].

\bibitem{Pomarol:1993mu}
A.~Pomarol and R.~Vega, {\it {Constraints on CP violation in the Higgs sector
  from the rho parameter}},  {\em Nucl. Phys.} {\bf B413} (1994) 3--15,
  [\href{http://xxx.lanl.gov/abs/hep-ph/9305272}{{\tt hep-ph/9305272}}].

\bibitem{Barger:1989fj}
V.~D. Barger, J.~L. Hewett, and R.~J.~N. Phillips, {\it {New constraints on the
  charged Higgs sector in two Higgs doublet models}},  {\em Phys. Rev.} {\bf
  D41} (1990) 3421.

\bibitem{Deshpande:1977rw}
N.~G. Deshpande and E.~Ma, {\it {Pattern of Symmetry Breaking with Two Higgs
  Doublets}},  {\em Phys. Rev.} {\bf D18} (1978) 2574.

\bibitem{Akeroyd:2000wc}
A.~G. Akeroyd, A.~Arhrib, and E.-M. Naimi, {\it {Note on tree-level unitarity
  in the general two Higgs doublet model}},  {\em Phys. Lett.} {\bf B490}
  (2000) 119--124, [\href{http://xxx.lanl.gov/abs/hep-ph/0006035}{{\tt
  hep-ph/0006035}}].

\bibitem{Ginzburg:2005dt}
I.~F. Ginzburg and I.~P. Ivanov, {\it {Tree-level unitarity constraints in the
  most general 2HDM}},  {\em Phys. Rev.} {\bf D72} (2005) 115010,
  [\href{http://xxx.lanl.gov/abs/hep-ph/0508020}{{\tt hep-ph/0508020}}].

\bibitem{Tobe:2002zj}
K.~Tobe and J.~D. Wells, {\it {Higgs boson mass limits in perturbative
  unification theories}},  {\em Phys. Rev.} {\bf D66} (2002) 013010,
  [\href{http://xxx.lanl.gov/abs/hep-ph/0204196}{{\tt hep-ph/0204196}}].

\bibitem{Peskin:1991sw}
M.~E. Peskin and T.~Takeuchi, {\it Estimation of oblique electroweak
  corrections},  {\em Phys. Rev.} {\bf D46} (1992) 381--409.

\bibitem{Grimus:2007if}
W.~Grimus, L.~Lavoura, O.~M. Ogreid, and P.~Osland, {\it {A precision
  constraint on multi-Higgs-doublet models}},  {\em J. Phys.} {\bf G35} (2008)
  075001, [\href{http://xxx.lanl.gov/abs/0711.4022}{{\tt 0711.4022}}].

\bibitem{Lytel:1980zh}
R.~Lytel, {\it Weak isospin breaking and higher order corrections},  {\em Phys.
  Rev.} {\bf D22} (1980) 505.

\bibitem{Haber:1993wf}
H.~E. Haber, {\it {Introductory low-energy supersymmetry}},
  \href{http://xxx.lanl.gov/abs/hep-ph/9306207}{{\tt hep-ph/9306207}}.

\bibitem{He:2001tp}
H.-J. He, N.~Polonsky, and S.-f. Su, {\it {Extra families, Higgs spectrum and
  oblique corrections}},  {\em Phys. Rev.} {\bf D64} (2001) 053004,
  [\href{http://xxx.lanl.gov/abs/hep-ph/0102144}{{\tt hep-ph/0102144}}].

\bibitem{lepewwg}
``{LEP Electroweak Working Group}.'' http://lepewwg.web.cern.ch/.

\bibitem{Coleman:1973jx}
S.~R. Coleman and E.~J. Weinberg, {\it {Radiative Corrections as the Origin of
  Spontaneous Symmetry Breaking}},  {\em Phys. Rev.} {\bf D7} (1973)
  1888--1910.

\bibitem{Barberio:2007cr}
{\bf Heavy Flavor Averaging Group (HFAG)} Collaboration, E.~Barberio {\em
  et~al.}, {\it {Averages of b-hadron properties at the end of 2006}},
  \href{http://xxx.lanl.gov/abs/arXiv:0704.3575 [hep-ex]}{{\tt arXiv:0704.3575
  [hep-ex]}}.

\bibitem{Grinstein:1987pu}
B.~Grinstein and M.~B. Wise, {\it {Weak Radiative B Meson Decay as a Probe of
  the Higgs Sector}},  {\em Phys. Lett.} {\bf B201} (1988) 274.

\bibitem{Hou:1987kf}
W.-S. Hou and R.~S. Willey, {\it {Effects of Charged Higgs Bosons on the
  Processes b $\to$ s Gamma, b $\to$ s g* and b $\to$ s Lepton+ Lepton-}},
  {\em Phys. Lett.} {\bf B202} (1988) 591.

\bibitem{Borzumati:1998tg}
F.~Borzumati and C.~Greub, {\it {2HDMs predictions for anti-B $\rightarrow$ X/s
  gamma in NLO {QCD}}},  {\em Phys. Rev.} {\bf D58} (1998) 074004,
  [\href{http://xxx.lanl.gov/abs/hep-ph/9802391}{{\tt hep-ph/9802391}}].

\bibitem{Misiak:2006zs}
M.~Misiak {\em et~al.}, {\it {The first estimate of B(anti-B $\rightarrow$ X/s
  gamma) at O(alpha(s)**2)}},  {\em Phys. Rev. Lett.} {\bf 98} (2007) 022002,
  [\href{http://xxx.lanl.gov/abs/hep-ph/0609232}{{\tt hep-ph/0609232}}].

\bibitem{Xiao:2003ya}
Z.-j. Xiao and L.~Guo, {\it {B0 anti-B0 mixing and B $\rightarrow$ X/s gamma
  decay in the third type 2HDM: Effects of NLO QCD contributions}},  {\em Phys.
  Rev.} {\bf D69} (2004) 014002,
  [\href{http://xxx.lanl.gov/abs/hep-ph/0309103}{{\tt hep-ph/0309103}}].

\bibitem{:2007ds}
{\bf BABAR} Collaboration, B.~Aubert {\em et~al.}, {\it {Observation of the
  Semileptonic Decays B $\rightarrow$ D* tau nubar and Evidence for B
  $\rightarrow$ D tau nubar}},  {\em Phys. Rev. Lett.} {\bf 100} (2008) 021801,
  [\href{http://xxx.lanl.gov/abs/arXiv:0709.1698 [hep-ex]}{{\tt arXiv:0709.1698
  [hep-ex]}}].

\bibitem{Nierste:2008qe}
U.~Nierste, S.~Trine, and S.~Westhoff, {\it {Charged-Higgs effects in a new B
  -> D tau nu differential decay distribution}},  {\em Phys. Rev.} {\bf D78}
  (2008) 015006, [\href{http://xxx.lanl.gov/abs/0801.4938}{{\tt 0801.4938}}].

\bibitem{Ikado:2006un}
K.~Ikado {\em et~al.}, {\it {Evidence of the purely leptonic decay B-
  $\rightarrow$ tau- anti- nu/tau}},  {\em Phys. Rev. Lett.} {\bf 97} (2006)
  251802, [\href{http://xxx.lanl.gov/abs/hep-ex/0604018}{{\tt
  hep-ex/0604018}}].

\bibitem{smith:2008}
C.~Smith, {\it {Minimal Flavour Violation}},  {\em Lectures notes} (2008).

\bibitem{:2007xj}
{\bf BABAR} Collaboration, B.~Aubert {\em et~al.}, {\it {A search for B+
  $\rightarrow$ tau+ nu with Hadronic B tags}},  {\em Phys. Rev.} {\bf D77}
  (2008) 011107, [\href{http://xxx.lanl.gov/abs/arXiv:0708.2260 [hep-ex]}{{\tt
  arXiv:0708.2260 [hep-ex]}}].

\bibitem{Geng:1988bq}
C.~Q. Geng and J.~N. Ng, {\it {Charged Higgs effect in B(d)0 - anti-B(d)0
  mixing, K $\to$ pi neutrino anti-neutrino decay and rare decays of B
  mesons}},  {\em Phys. Rev.} {\bf D38} (1988) 2857.

\bibitem{Ball:2006xx}
P.~Ball and R.~Fleischer, {\it {Probing new physics through B mixing: Status,
  benchmarks and prospects}},  {\em Eur. Phys. J.} {\bf C48} (2006) 413--426,
  [\href{http://xxx.lanl.gov/abs/hep-ph/0604249}{{\tt hep-ph/0604249}}].

\bibitem{Yao:2006px}
{\bf Particle Data Group} Collaboration, W.~M. Yao {\em et~al.}, {\it Review of
  particle physics},  {\em J. Phys.} {\bf G33} (2006) 1--1232.

\bibitem{Urban:1997gw}
J.~Urban, F.~Krauss, U.~Jentschura, and G.~Soff, {\it {Next-to-leading order
  QCD corrections for the B0 anti-B0 mixing with an extended Higgs sector}},
  {\em Nucl. Phys.} {\bf B523} (1998) 40--58,
  [\href{http://xxx.lanl.gov/abs/hep-ph/9710245}{{\tt hep-ph/9710245}}].

\bibitem{Logan:1999if}
H.~E. Logan, {\it {Radiative corrections to the Z b anti-b vertex and
  constraints on extended Higgs sectors}},
  \href{http://xxx.lanl.gov/abs/hep-ph/9906332}{{\tt hep-ph/9906332}}.

\bibitem{Hagiwara:2006jt}
K.~Hagiwara, A.~D. Martin, D.~Nomura, and T.~Teubner, {\it {Improved
  predictions for g-2 of the muon and $\alpha_{\rm QED}(M_Z^2)$}},  {\em Phys.
  Lett.} {\bf B649} (2007) 173--179,
  [\href{http://xxx.lanl.gov/abs/hep-ph/0611102}{{\tt hep-ph/0611102}}].

\bibitem{Bennett:2006fi}
{\bf Muon G-2} Collaboration, G.~W. Bennett {\em et~al.}, {\it {Final report of
  the muon E821 anomalous magnetic moment measurement at BNL}},  {\em Phys.
  Rev.} {\bf D73} (2006) 072003,
  [\href{http://xxx.lanl.gov/abs/hep-ex/0602035}{{\tt hep-ex/0602035}}].

\bibitem{Krawczyk:2001pe}
M.~Krawczyk, {\it {The new (g-2)(mu) measurement and limits on the light Higgs
  bosons in 2HDM. II}},  \href{http://xxx.lanl.gov/abs/hep-ph/0103223}{{\tt
  hep-ph/0103223}}.

\bibitem{Barate:2003sz}
{\bf LEP Working Group for Higgs boson searches} Collaboration, R.~Barate {\em
  et~al.}, {\it {Search for the standard model Higgs boson at LEP}},  {\em
  Phys. Lett.} {\bf B565} (2003) 61--75,
  [\href{http://xxx.lanl.gov/abs/hep-ex/0306033}{{\tt hep-ex/0306033}}].

\bibitem{Accomando:2006ga}
E.~Accomando {\em et~al.}, {\it {Workshop on CP studies and non-standard Higgs
  physics}},  \href{http://xxx.lanl.gov/abs/hep-ph/0608079}{{\tt
  hep-ph/0608079}}.

\bibitem{Schael:2006cr}
{\bf LEP Working Group for Higgs boson searches} Collaboration, S.~Schael {\em
  et~al.}, {\it {Search for neutral MSSM Higgs bosons at LEP}},  {\em Eur.
  Phys. J.} {\bf C47} (2006) 547--587,
  [\href{http://xxx.lanl.gov/abs/hep-ex/0602042}{{\tt hep-ex/0602042}}].

\bibitem{Abdallah:2004wy}
{\bf DELPHI} Collaboration, J.~Abdallah {\em et~al.}, {\it {Searches for
  neutral Higgs bosons in extended models}},  {\em Eur. Phys. J.} {\bf C38}
  (2004) 1--28, [\href{http://xxx.lanl.gov/abs/hep-ex/0410017}{{\tt
  hep-ex/0410017}}].

\bibitem{Krawczyk:1998kk}
M.~Krawczyk, J.~Zochowski, and P.~Mattig, {\it {Process Z $\rightarrow$ h(A) +
  gamma in the 2HDM and the experimental constraints from LEP}},  {\em Eur.
  Phys. J.} {\bf C8} (1999) 495--505,
  [\href{http://xxx.lanl.gov/abs/hep-ph/9811256}{{\tt hep-ph/9811256}}].

\bibitem{Abdallah:2003wd}
{\bf DELPHI} Collaboration, J.~Abdallah {\em et~al.}, {\it {Search for charged
  Higgs bosons at LEP in general two Higgs doublet models}},  {\em Eur. Phys.
  J.} {\bf C34} (2004) 399--418,
  [\href{http://xxx.lanl.gov/abs/hep-ex/0404012}{{\tt hep-ex/0404012}}].

\bibitem{:2008be}
{\bf OPAL} Collaboration, G.~Abbiendi {\em et~al.}, {\it {Search for Charged
  Higgs Bosons in e +e- Collisions at sqrts(s) = 189-209 GeV}},
  \href{http://xxx.lanl.gov/abs/0812.0267}{{\tt 0812.0267}}.

\bibitem{Akeroyd:1998dt}
A.~G. Akeroyd, {\it {Three-body decays of Higgs bosons at LEP2 and application
  to a hidden fermiophobic Higgs}},  {\em Nucl. Phys.} {\bf B544} (1999)
  557--575, [\href{http://xxx.lanl.gov/abs/hep-ph/9806337}{{\tt
  hep-ph/9806337}}].

\bibitem{Tev2009Higgs}
{\bf CDF/D0} Collaboration, {\it {Combined CDF and D0 Upper Limits on Standard
  Model Higgs-Boson Production with up to 4.2 fb$^{-1}$ of Data}}, .
  FERMILAB-PUB-09-060-E.

\bibitem{Stelzer:2006sp}
T.~Stelzer, S.~Wiesenfeldt, and S.~Willenbrock, {\it {Higgs at the Tevatron in
  extended supersymmetric models}},  {\em Phys. Rev.} {\bf D75} (2007) 077701,
  [\href{http://xxx.lanl.gov/abs/hep-ph/0611242}{{\tt hep-ph/0611242}}].

\bibitem{Cheung:2007sva}
K.~Cheung, J.~Song, and Q.-S. Yan, {\it {Role of $h \to \eta \eta$ in
  Intermediate-Mass Higgs Boson Searches at the Large Hadron Collider}},  {\em
  Phys. Rev. Lett.} {\bf 99} (2007) 031801,
  [\href{http://xxx.lanl.gov/abs/hep-ph/0703149}{{\tt hep-ph/0703149}}].

\bibitem{Djouadi:2008uw}
A.~Djouadi {\em et~al.}, {\it {Benchmark scenarios for the NMSSM}},  {\em JHEP}
  {\bf 07} (2008) 002, [\href{http://xxx.lanl.gov/abs/0801.4321}{{\tt
  0801.4321}}].

\bibitem{Carena:2007jk}
M.~Carena, T.~Han, G.-Y. Huang, and C.~E.~M. Wagner, {\it {Higgs Signal for h
  $\to$ aa at Hadron Colliders}},  {\em JHEP} {\bf 04} (2008) 092,
  [\href{http://xxx.lanl.gov/abs/0712.2466}{{\tt 0712.2466}}].

\bibitem{Roco:2001}
M.~Roco {\em FERMILAB-Conf} (2001), no.~00/203-E.

\bibitem{CDF:2007}
{CDF Collaboration}, {\it {Search for Neutral MSSM Higgs Bosons Decaying to Tau
  Pairs with $1.8 fb^{-1}$ of Data}},  {\em CDF Note} (2007), no.~9071.

\bibitem{Akeroyd:2003jp}
A.~G. Akeroyd, {\it {Searching for a very light Higgs boson at the Tevatron}},
  {\em Phys. Rev.} {\bf D68} (2003) 077701,
  [\href{http://xxx.lanl.gov/abs/hep-ph/0306045}{{\tt hep-ph/0306045}}].

\bibitem{Akeroyd:2007yj}
A.~G. Akeroyd, A.~Arhrib, and Q.-S. Yan, {\it {Charged Higgs bosons in the
  Next-to MSSM (NMSSM)}},  {\em Eur. Phys. J.} {\bf C55} (2008) 653--665,
  [\href{http://xxx.lanl.gov/abs/0712.3933}{{\tt 0712.3933}}].

\bibitem{CDF2008Charged}
{\bf CDF} Collaboration, {\it {A search for charged Higgs in lepton + jets
  $\overline{t}t$ events using 2.2 fb$^{-1}$ of CDF data}}, . CDF note 9322.

\bibitem{Alwall:2007st}
J.~Alwall {\em et~al.}, {\it {MadGraph/MadEvent v4: The New Web Generation}},
  {\em JHEP} {\bf 09} (2007) 028,
  [\href{http://xxx.lanl.gov/abs/0706.2334}{{\tt 0706.2334}}].

\bibitem{Alwall:2008pm}
J.~Alwall {\em et~al.}, {\it {New Developments in MadGraph/MadEvent}},  {\em
  AIP Conf. Proc.} {\bf 1078} (2009) 84--89,
  [\href{http://xxx.lanl.gov/abs/0809.2410}{{\tt 0809.2410}}].

\bibitem{Pumplin:2002vw}
J.~Pumplin {\em et~al.}, {\it {New generation of parton distributions with
  uncertainties from global QCD analysis}},  {\em JHEP} {\bf 07} (2002) 012,
  [\href{http://xxx.lanl.gov/abs/hep-ph/0201195}{{\tt hep-ph/0201195}}].

\bibitem{Sjostrand:2006za}
T.~Sjostrand, S.~Mrenna, and P.~Skands, {\it {PYTHIA 6.4 physics and manual}},
  {\em JHEP} {\bf 05} (2006) 026,
  [\href{http://xxx.lanl.gov/abs/hep-ph/0603175}{{\tt hep-ph/0603175}}].

\bibitem{XavSev}
S.~Ovyn, X.~Rouby, and V.~Lemaitre, {\it {Delphes, a framework for fast
  simulation of a generic collider experiment}},
  \href{http://xxx.lanl.gov/abs/0903.2225}{{\tt 0903.2225}}.

\bibitem{Cacciari:2005hq}
M.~Cacciari and G.~P. Salam, {\it {Dispelling the $N^3$ myth for the k(t)
  jet-finder}},  {\em Phys. Lett.} {\bf B641} (2006) 57--61,
  [\href{http://xxx.lanl.gov/abs/hep-ph/0512210}{{\tt hep-ph/0512210}}].

\bibitem{Djouadi:2005gj}
A.~Djouadi, {\it {The anatomy of electro-weak symmetry breaking. II: The Higgs
  bosons in the minimal supersymmetric model}},
  \href{http://xxx.lanl.gov/abs/hep-ph/0503173}{{\tt hep-ph/0503173}}.

\bibitem{Harlander:2003ai}
R.~V. Harlander and W.~B. Kilgore, {\it {Higgs boson production in bottom quark
  fusion at next-to- next-to-leading order}},  {\em Phys. Rev.} {\bf D68}
  (2003) 013001, [\href{http://xxx.lanl.gov/abs/hep-ph/0304035}{{\tt
  hep-ph/0304035}}].

\bibitem{Alwall:2008qv}
J.~Alwall, S.~de~Visscher, and F.~Maltoni, {\it {QCD radiation in the
  production of heavy colored particles at the LHC}},  {\em JHEP} {\bf 02}
  (2009) 017, [\href{http://xxx.lanl.gov/abs/0810.5350}{{\tt 0810.5350}}].

\bibitem{Plehn:2002vy}
T.~Plehn, {\it {Charged Higgs boson production in bottom gluon fusion}},  {\em
  Phys. Rev.} {\bf D67} (2003) 014018,
  [\href{http://xxx.lanl.gov/abs/hep-ph/0206121}{{\tt hep-ph/0206121}}].

\bibitem{:1999fq}
{\it {ATLAS: Detector and physics performance technical design report. Volume
  1}}, . CERN-LHCC-99-14.

\bibitem{Ball:2007zza}
{\bf CMS} Collaboration, G.~L. Bayatian {\em et~al.}, {\it {CMS technical
  design report, volume II: Physics performance}},  {\em J. Phys.} {\bf G34}
  (2007) 995--1579.

\bibitem{Catani:2003zt}
S.~Catani, D.~de~Florian, M.~Grazzini, and P.~Nason, {\it {Soft-gluon
  resummation for Higgs boson production at hadron colliders}},  {\em JHEP}
  {\bf 07} (2003) 028, [\href{http://xxx.lanl.gov/abs/hep-ph/0306211}{{\tt
  hep-ph/0306211}}].

\bibitem{Butterworth:2008sd}
J.~M. Butterworth, A.~R. Davison, M.~Rubin, and G.~P. Salam, {\it {Jet
  substructure as a new Higgs search channel at the LHC}},  {\em AIP Conf.
  Proc.} {\bf 1078} (2009) 189--191,
  [\href{http://xxx.lanl.gov/abs/0809.2530}{{\tt 0809.2530}}].

\end{thebibliography}\endgroup

\end{document}